\documentclass[12pt,prd,showpacs,tightenlines,nofootinbib]{revtex4}
\usepackage{bm}
\usepackage{graphics}
\usepackage{rotating}
\usepackage{epsfig}
\begin{document}
\begin{flushright}{HU-EP-10/30}\end{flushright}
\title{
Rare semileptonic decays of $B$ and $B_c$ mesons
in the relativistic quark model}
\author{D. Ebert$^{1}$, R. N. Faustov$^{1,2}$  and V. O. Galkin$^{1,2}$}
\affiliation{
$^1$ Institut f\"ur Physik, Humboldt--Universit\"at zu Berlin,
Newtonstr. 15, D-12489  Berlin, Germany\\
$^2$ Dorodnicyn Computing Centre, Russian Academy of Sciences,
  Vavilov Str. 40, 119991 Moscow, Russia}

\begin{abstract}
Rare semileptonic decays of $B$ and $B_c$ mesons are investigated in
the framework of the QCD-motivated relativistic quark model based on
the quasipotential approach. Form  factors parametrizing the matrix
elements of the weak transitions between corresponding meson
states are calculated with the complete account of the relativistic
effects including contributions of intermediate negative energy states
and relativistic transformations of the meson wave functions. The
momentum transfer dependence of the form factors is reliably
determined in the whole accessible kinematical range. On this basis the
total and differential branching fractions of the $B\to
K^{(*)}l^+l^-(\nu\bar\nu)$ and $B_c\to D_s^{(*)}l^+l^-(\nu\bar\nu)$,
$B_c\to D^{(*)}l^+l^-(\nu\bar\nu)$ decays as well as the
longitudinal polarization fractions $F_L$ of the  final vector meson and
the muon forward-backward 
asymmetries $A_{FB}$ are calculated. Good agreement of the obtained
results with the recent detailed experimental data
on the $B\to K^{(*)}\mu^+\mu^-$ decays  from Belle and CDF is found. Predictions for
the rare semileptonic decays of the $B_c$ mesons are given.

\end{abstract}

\pacs{13.20.He, 12.39.Ki}

\maketitle

\section{Introduction}
\label{sec:intr}

The investigation of the rare weak $B$ and $B_c$ meson decays
represents a very interesting and important problem. Such
decays are governed by the flavour-changing neutral currents, which
are forbidden at tree level in the standard model (SM) and first appear at
one-loop. Therefore, such decays are very sensitive to the
contributions of new intermediate particles and interactions,
predicted in numerous extensions of the SM (see e.g. \cite{ehmrr,absw}
and references therein). Notwithstanding the fact that such decays
have very small 
branching ratios, comparison of existing theoretical and experimental
results for the rare semileptonic and radiative $B\to K^{(*)}$ decays
already provides one of the most rigid 
constraints on different new physics scenarios \cite{hn}.    

 The theoretical analysis of the rare weak $B$ decays is based on the
electroweak effective Hamiltonian, which is obtained by integrating
out the heavy degrees of freedom (electroweak bosons and top
quark) \cite{hn}. The QCD corrections to these 
processes due to hard gluon exchanges turn out to be important and
require to resum large logarithms, which is done with the help of
renormalization group methods. The operator product expansion allows
to separate the short-distance part in the $B$ meson decay amplitudes,
which is described by the Wilson coefficients and can be calculated
perturbatively, from the long-distance part contained in the operator matrix
elements between initial and final meson states.  For the investigation
of the exclusive decay rates one needs to apply nonperturbative
methods to calculate these hadronic matrix elements which are usually
parametrized in terms of covariant form factors. Clearly, such calculation
is model dependent. In order to reduce model dependence, methods, based
on the heavy quark and large energy expansions, have been
developed. They employ the new symmetries which arise in heavy quark and
large energy limits and permit to significantly reduce the number of independent
form factors \cite{clopr}. Such methods allow a perturbative calculation of QCD
corrections to the factorization approximation and thus are now popular in the
literature \cite{scet}. However, the important $\Lambda_{QCD}/m_b$
corrections cannot be systematically taken into account in such an
approach. 

Rare $B\to K^{(*)}$ transitions are the most studied ones both theoretically and
experimentally \cite{hn}. Recently, detailed experimental data on differential
branching fractions, angular distributions  and asymmetries in the rare $B\to
K^{(*)}\mu^+\mu^-$ decays became available both from $B$ factories and
Tevatron \cite{babarbk,bellebk,cdfbk,cdfbk1}. The measured values are
at present consistent with the predictions of the SM
within experimental and theoretical  uncertainties. Significantly
better statistics on the rare $B$ decays is expected form LHC experiments
(especially from LHCb) which will allow precision tests of the
SM and can probably reveal signals of new physics \cite{LHC}. It is
expected that the $B_c$ mesons will be copiously produced at LHC, making
possible the experimental study of their weak rare decays. Such decays
received significantly less attention in the literature. The $B_c\to
D_s^{(*)}l^+l^-$ and $B_c\to D^{(*)}l^+l^-$ were previously
investigated using the relativistic constituent quark
model \cite{fgikl}, light-front quark model \cite{ghl,choi} and three-point QCD sum
rules \cite{azizi}.       

In this paper we study the rare weak $B$ and $B_c$ decays in the framework
of the QCD-motivated relativistic quark model. Our model was
previously successfully applied for the investigation of various
electroweak properties of heavy and light hadrons. Semileptonic decay rates
of the $B$ \cite{bdecays} and $B_c$ \cite{bcjpsi,bcbs} mesons as well as rare
radiative decays of the $B$ \cite{brare} were calculated in agreement with
available experimental data. For this purpose, effective methods of the
calculation of electroweak matrix elements between meson states
with a consistent account of relativistic effects were
developed. They allow to reliably determine the form factor dependence
on the momentum transfer in the whole accessible kinematical
range. The form factors are expressed as overlap integrals of the
meson wave functions, which were obtained in the corresponding
calculations of the mass spectra \cite{mass,hlm}. It is important to note that we specially checked
\cite{fg,ffhm} the fulfillment of the model-independent symmetry
relations among form factors arising in the heavy quark and large
energy limits. Here we apply these methods to the calculation of the
form factors of the rare $B\to K^{(*)}$ and $B_c\to
D_s^{(*)}(D^{(*)})$ transitions and on this basis determine 
branching fractions and differential distributions of these decays.  

The paper is organized as follows. The relevant effective weak Hamiltonian for the rare
$B$ and $B_c$ decays is briefly discussed in Sec.~\ref{sec:eh}. In
Sec.~\ref{rqm} we give an outline of our relativistic quark model. Then
in Sec.~\ref{mml}   we 
discuss the relativistic calculation of the hadronic matrix element of
the weak current between meson states in the quasipotential
approach. Special attention is devoted to the contributions of 
negative energy states and the relativistic transformation of the wave
functions from the rest to the moving reference frame. Form factors of
the rare semileptonic $B\to K^{(*)}$ and $B_c\to
D_s^{(*)}(D^{(*)})$ decays are calculated in Sec.~\ref{ffr}. These
form factors are used in Sec.~\ref{sec:rd} for the calculation of the
total and differential rare decay branching fractions. First we give
 the necessary formulas and then present our numerical results. These are then
confronted with available experimental data and predictions of other
approaches.  Finally, Sec.~\ref{sec:concl} contains our conclusions. 
Expressions for the tensor form factors of the rare $B$ and $B_c$
meson decays in terms of the overlap integrals of meson wave functions
are given in the Appendix.   

\section{Effective Hamiltonian for the rare $B$ and $B_c$ meson decays}
\label{sec:eh}
 
The usual approach to the description of rare $B$ decays is based on the
low-energy effective Hamiltonian, obtained by integrating out the
heavy degrees of freedom (the top quark and $W$ bosons) of the
SM. The operator product 
expansion separates the short-distance contributions, which are
contained in the Wilson coefficients and can be calculated perturbatively,
from the long-distance contributions contained in the matrix elements
of the local operators. The calculation of such matrix elements
requires the application of nonperturbative methods. 

The effective Hamiltonians for $b\to f l^+l^-$ and $b\to f \nu\bar \nu$
transitions ($f=s$ or $d$), renormalized at a scale $\mu\approx m_b$,
are given by \cite{bhi}       
\begin{eqnarray}
  \label{eq:heff}
  {\cal H}_{\rm eff}^{l^+l^-} &=&-\frac{4G_F}{\sqrt{2}}V_{tf}^*V_{tb}\sum_{i=1}^{10}C_i{\cal
      O}_i,\cr
 {\cal H}_{\rm eff}^{\nu\bar\nu}&=&-\frac{4G_F}{\sqrt{2}}V_{tf}^*V_{tb}C^\nu_L {\cal O}^\nu_L,
\end{eqnarray}
where $G_F$ is the Fermi constant, $V_{tj}$ are
Cabibbo-Kobayashi-Maskawa matrix elements, $C_i$ are the Wilson coefficients
and ${\cal O}_i$ are the standard model  operator basis which can be
found e.g. in \cite{bbl}.  The most important operators  for the $b\to f
l^+l^-$ transitions are the following 
\begin{eqnarray}
  \label{eq:op}
  {\cal O}_7&=&\frac{e}{32\pi^2}m_b(\bar
  f\sigma_{\mu\nu}(1+\gamma_5)b)F^{\mu\nu},\cr\cr
{\cal O}_9&=&\frac{e^2}{32\pi^2}(\bar
f\gamma_{\mu}(1-\gamma_5)b)(\bar l\gamma^\mu l),\cr\cr
{\cal O}_{10}&=&\frac{e^2}{32\pi^2}(\bar
f\gamma_{\mu}(1-\gamma_5)b)(\bar l\gamma^\mu\gamma_5 l),
\end{eqnarray}
with $F_{\mu\nu}$ being the electromagnetic field strength tensor,
and for the $b\to f \nu\bar \nu$ transitions we have
\begin{equation}
  \label{eq:ocl}
  {\cal O}_L^\nu=\frac{e^2}{32\pi^2}(\bar
f\gamma_{\mu}(1-\gamma_5)b)(\bar \nu\gamma^\mu(1-\gamma_5)\nu).
\end{equation}

The resulting structure of the free quark decay amplitude has the form:
\begin{eqnarray}
  \label{eq:wda}
  M(b\to fl^+l^-)&=& \frac{G_F}{\sqrt{2}}\frac{\alpha}{2\pi}
  V_{tf}^*V_{tb}\Bigl[C_9^{\rm eff}(\bar f\gamma_{\mu}(1-\gamma_5)b)
 (\bar l\gamma^\mu l)+C_{10}(\bar
f\gamma_{\mu}(1-\gamma_5)b)(\bar l\gamma^\mu\gamma_5 l)\cr
&&-\frac{2m_b}{q^2}C_7^{\rm eff}(\bar
  f\sigma_{\mu\nu}q^\nu(1+\gamma_5)b)(\bar l\gamma^\mu l)\Bigr],\cr
M(b\to f\nu\bar\nu)&=& \frac{G_F}{\sqrt{2}}\frac{\alpha}{2\pi}
  V_{tf}^*V_{tb}C^\nu_L(\bar
f\gamma_{\mu}(1-\gamma_5)b)(\bar \nu\gamma^\mu(1-\gamma_5)\nu),
\end{eqnarray}
where $\alpha$ is the fine structure constant.

The effective Wilson coefficient $C_7^{\rm eff}$ is given \cite{abhh} by $C_7^{\rm eff}=C_7-C_5/3-C_6$, while
$C_9^{\rm eff}$ accounts for  both perturbative and certain
long-distance contributions from the matrix elements of
four-quark operators ${\cal O}_{1, \dots, 6}$. The long-distance
(nonperturbative) effects arise from the $c\bar c$ resonance
contributions from $J/\psi, \psi'\dots$ and are usually assumed to have
a phenomenological Breit-Wigner structure. Therefore $C_9^{\rm eff}$
reads as follows \cite{abhh,fgikl,mns}
\begin{equation}
  \label{eq:ceff9}
  C_9^{\rm eff}=C_9+{\cal Y}_{\rm pert}(q^2)+{\cal Y}_{\rm BW}(q^2).
\end{equation}
Here the perturbative part is given by
\begin{eqnarray}
  \label{eq:ypert}
{\cal Y}_{\rm pert}(q^2)&=&h\left(\frac{m_c}{m_b},\frac{q^2}{m_b^2}\right)(3C_1+C_2+3C_3+C_4+3C_5+C_6)\cr
&&-
\frac12 h\left(1,\frac{q^2}{m_b^2}\right)(4C_3+4C_4+3C_5+C_6)\cr
&&-\frac12 h\left(0,\frac{q^2}{m_b^2}\right)(C_3+3C_4)+\frac29(3C_3+C_4+3C_5+C_6),  
\end{eqnarray}
and the $c\bar c$ resonance part reads
\begin{equation}
  \label{eq:ybw}
{\cal Y}_{\rm BW}(q^2)=\frac{3\pi}{\alpha^2} \sum_{V_i=J/\psi,\psi'}\frac{\Gamma(V_i\to l^+l^-)M_{V_i} }{M_{V_i}^2-q^2-iM_{V_i}\Gamma_{V_i}},
\end{equation}
and $q^2$ is the four-momentum squared of the lepton pair, $m_{b,c}$ are the
masses of the $b$ and $c$ quarks.
The explicit form of the function $h(m_c/m_b,q^2/m^2)$ \cite{bm}  and the
values of Wilson 
coefficients $C_{1,\dots,10}$ are given in Refs.~\cite{abhh,fgikl}. 

For the application of the above expressions to the description of the
exclusive rare semileptonic decays of the $B$ 
and $B_c$ mesons it is necessary to calculate the matrix elements of
the operators $\bar f\gamma_{\mu}(1-\gamma_5)b$ and $\bar
f\sigma_{\mu\nu}q^\nu(1-\gamma_5)b$ between initial and final hadron
states. Such calculation requires the application of nonperturbative
approaches. In this paper we use the relativistic quark model based on the
quasipotential approach for these investigations.

\section{Relativistic quark model}  
\label{rqm}

In the quasipotential approach a meson is described as a bound
quark-antiquark state with a wave function satisfying the
quasipotential equation of the Schr\"odinger type 
\begin{equation}
\label{quas}
{\left(\frac{b^2(M)}{2\mu_{R}}-\frac{{\bf
p}^2}{2\mu_{R}}\right)\Psi_{M}({\bf p})} =\int\frac{d^3 q}{(2\pi)^3}
 V({\bf p,q};M)\Psi_{M}({\bf q}),
\end{equation}
where the relativistic reduced mass is
\begin{equation}
\mu_{R}=\frac{E_1E_2}{E_1+E_2}=\frac{M^4-(m^2_1-m^2_2)^2}{4M^3},
\end{equation}
and $E_1$, $E_2$ are the center of mass energies on mass shell given by
\begin{equation}
\label{ee}
E_1=\frac{M^2-m_2^2+m_1^2}{2M}, \quad E_2=\frac{M^2-m_1^2+m_2^2}{2M}.
\end{equation}
Here $M=E_1+E_2$ is the meson mass, $m_{1,2}$ are the quark masses,
and ${\bf p}$ is their relative momentum.  
In the center of mass system the relative momentum squared on mass shell 
reads
\begin{equation}
{b^2(M) }
=\frac{[M^2-(m_1+m_2)^2][M^2-(m_1-m_2)^2]}{4M^2}.
\end{equation}

The kernel 
$V({\bf p,q};M)$ in Eq.~(\ref{quas}) is the quasipotential operator of
the quark-antiquark interaction. It is constructed with the help of the
off-mass-shell scattering amplitude, projected onto the positive
energy states. 
Constructing the quasipotential of the quark-antiquark interaction, 
we have assumed that the effective
interaction is the sum of the usual one-gluon exchange term with the mixture
of long-range vector and scalar linear confining potentials, where
the vector confining potential
contains the Pauli interaction. The quasipotential is then defined by
\cite{mass}
  \begin{equation}
\label{qpot}
V({\bf p,q};M)=\bar{u}_1(p)\bar{u}_2(-p){\mathcal V}({\bf p}, {\bf
q};M)u_1(q)u_2(-q),
\end{equation}
with
$${\mathcal V}({\bf p},{\bf q};M)=\frac{4}{3}\alpha_sD_{ \mu\nu}({\bf
k})\gamma_1^{\mu}\gamma_2^{\nu}
+V^V_{\rm conf}({\bf k})\Gamma_1^{\mu}
\Gamma_{2;\mu}+V^S_{\rm conf}({\bf k}),$$
where $\alpha_s$ is the QCD coupling constant, $D_{\mu\nu}$ is the
gluon propagator in the Coulomb gauge
\begin{equation}
D^{00}({\bf k})=-\frac{4\pi}{{\bf k}^2}, \quad D^{ij}({\bf k})=
-\frac{4\pi}{k^2}\left(\delta^{ij}-\frac{k^ik^j}{{\bf k}^2}\right),
\quad D^{0i}=D^{i0}=0,
\end{equation}
and ${\bf k=p-q}$. Here $\gamma_{\mu}$ and $u(p)$ are 
the Dirac matrices and spinors
\begin{equation}
\label{spinor}
u^\lambda({p})=\sqrt{\frac{\epsilon(p)+m}{2\epsilon(p)}}
\left(
\begin{array}{c}1\cr {\displaystyle\frac{\bm{\sigma}
      {\bf  p}}{\epsilon(p)+m}}
\end{array}\right)\chi^\lambda,
\end{equation}
where  $\bm{\sigma}$   and $\chi^\lambda$
are Pauli matrices and spinors and $\epsilon(p)=\sqrt{{\bf p}^2+m^2}$.
The effective long-range vector vertex is
given by
\begin{equation}
\label{kappa}
\Gamma_{\mu}({\bf k})=\gamma_{\mu}+
\frac{i\kappa}{2m}\sigma_{\mu\nu}k^{\nu},
\end{equation}
where $\kappa$ is the Pauli interaction constant characterizing the
long-range anomalous chromomagnetic moment of quarks. Vector and
scalar confining potentials in the nonrelativistic limit reduce to
\begin{eqnarray}
\label{vlin}
V_V(r)&=&(1-\varepsilon)(Ar+B),\nonumber\\ 
V_S(r)& =&\varepsilon (Ar+B),
\end{eqnarray}
reproducing 
\begin{equation}
\label{nr}
V_{\rm conf}(r)=V_S(r)+V_V(r)=Ar+B,
\end{equation}
where $\varepsilon$ is the mixing coefficient. 

The expression for the quasipotential of the heavy quarkonia,
expanded in $v^2/c^2$  can be found in Ref.~\cite{mass}. The
quasipotential for the heavy quark interaction with a light antiquark
without employing the nonrelativistic ($v/c$)  expansion for the light quark
is given in Ref.~\cite{hlm}.  All the parameters of
our model like quark masses, parameters of the linear confining potential
$A$ and $B$, mixing coefficient $\varepsilon$ and anomalous
chromomagnetic quark moment $\kappa$ are fixed from the analysis of
heavy quarkonium masses and radiative
decays. The quark masses
$m_b=4.88$ GeV, $m_c=1.55$ GeV, $m_s=0.5$ GeV, $m_{u,d}=0.33$ GeV and
the parameters of the linear potential $A=0.18$ GeV$^2$ and $B=-0.30$ GeV
have the values inherent for quark models.  The value of the mixing
coefficient of vector and scalar confining potentials $\varepsilon=-1$
has been determined from the consideration of the heavy quark expansion
for the semileptonic $B\to D$ decays
\cite{fg} and charmonium radiative decays \cite{mass}.
Finally, the universal Pauli interaction constant $\kappa=-1$ has been
fixed from the analysis of the fine splitting of heavy quarkonia ${
}^3P_J$- states \cite{mass} and  the heavy quark expansion for semileptonic
decays of heavy mesons \cite{fg} and baryons \cite{sbar}. Note that the 
long-range  magnetic contribution to the potential in our model
is proportional to $(1+\kappa)$ and thus vanishes for the 
chosen value of $\kappa=-1$ in accordance with the flux tube model.

\section{Matrix elements of the effective weak current operators for
  $\bm{\lowercase{b\to s,d}}$ transitions} \label{mml}

In order to calculate the exclusive rare semileptonic decay rate of the
$B$  ($B_c$) meson, it is necessary to determine the corresponding hadronic matrix
element of the  weak operators (\ref{eq:op}), (\ref{eq:ocl}) between meson states.
In the quasipotential approach,  the matrix element of the hadronic
weak current operator 
$J^W_\mu$, between a $B$  ($B_c$) meson with mass $M_{B}$ and
four-momentum $p_{B}$ and a final meson $F$ ($F=K^{(*)}$ or $D_s^{(*)}$ and
$D^{(*)}$) with mass $M_F$  and four-momentum $p_F$ takes the form \cite{f} 
\begin{equation}\label{mxet} 
\langle F(p_F) \vert J^W_\mu \vert B(p_{B})\rangle
=\int \frac{d^3p\, d^3q}{(2\pi )^6} \bar \Psi_{F\,{\bf p}_F}({\bf
p})\Gamma _\mu ({\bf p},{\bf q})\Psi_{B\,{\bf p}_{B}}({\bf q}),
\end{equation}
where $\Gamma _\mu ({\bf p},{\bf
q})$ is the two-particle vertex function and  
$\Psi_{M\,{\bf p}_M}$ are the
meson ($M=B,F)$ wave functions projected onto the positive energy states of
quarks and boosted to the moving reference frame with three-momentum ${\bf p}_M$.
\begin{figure}
  \centering
  \includegraphics{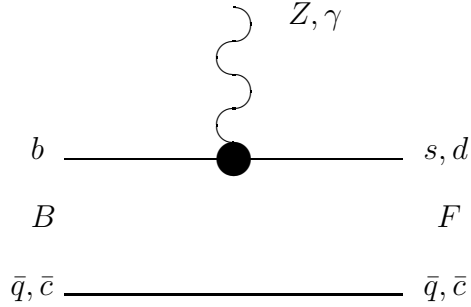}
\caption{Lowest order vertex function $\Gamma^{(1)}$
contributing to the current matrix element (\ref{mxet}). \label{d1}}
\end{figure}

\begin{figure}
  \centering
  \includegraphics{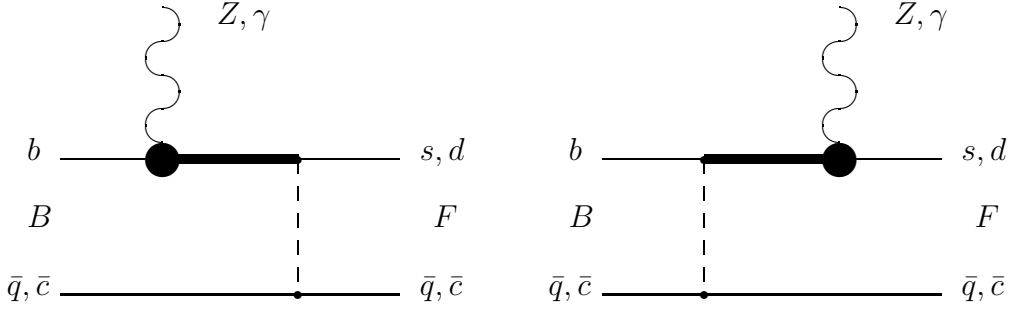}
\caption{ Vertex function $\Gamma^{(2)}$
taking the quark interaction into account. Dashed lines correspond  
to the effective potential ${\cal V}$ in 
(\ref{qpot}). Bold lines denote the negative-energy part of the quark
propagator. \label{d2}}
\end{figure}

 The contributions to $\Gamma$ come from Figs.~\ref{d1} and \ref{d2}. 
The leading order vertex function $\Gamma^{(1)}$ corresponds to the impulse
approximation, while the vertex function $\Gamma^{(2)}$ accounts for
contributions of the negative-energy states. Note that the form of the
relativistic corrections resulting from the vertex function
$\Gamma^{(2)}$ is explicitly dependent on the Lorentz structure of the
quark-antiquark interaction. In the leading order of the  the heavy
quark ($m_{b,c}\to \infty$) and large energy expansions for $B\to F$ transitions, 
only $\Gamma^{(1)}$ contributes, while $\Gamma^{(2)}$  
contributes already at the subleading order. 
The vertex functions are determined by
\begin{equation} \label{gamma1}
\Gamma_\mu^{(1)}({\bf
p},{\bf q})=\bar u_{f}(p_f){\cal G}_\mu u_b(q_b)
(2\pi)^3\delta({\bf p}_q-{\bf
q}_q),\end{equation}
and
\begin{eqnarray}\label{gamma2} 
\Gamma_\mu^{(2)}({\bf
p},{\bf q})&=&\bar u_{f}(p_f)\bar u_q(p_q) \Bigl\{{\cal G}_{1\mu}
\frac{\Lambda_b^{(-)}(
k)}{\epsilon_b(k)+\epsilon_b(p_q)}\gamma_1^0
{\cal V}({\bf p}_q-{\bf
q}_q)\nonumber \\ 
& &+{\cal V}({\bf p}_q-{\bf
q}_q)\frac{\Lambda_{f}^{(-)}(k')}{ \epsilon_{f}(k')+
\epsilon_{f}(q_f)}\gamma_1^0 {\cal G}_{1\mu}\Bigr\}u_b(q_b)
u_q(q_q),\end{eqnarray}
where ${\cal G}_\mu=\gamma_\mu(1-\gamma_5)$ for the (axial) vector weak current and
 ${\cal G}_\mu=\sigma_{\mu\nu}q^\nu(1+\gamma_5)$ for the (pseudo) tensor
 current; the subscripts $f$ and $q$ denote the final active $s,d$ and
 the spectator $u,d,c$ quarks, respectively; the superscripts ``(1)" and ``(2)" correspond to Figs.~\ref{d1} and
\ref{d2},  ${\bf k}={\bf p}_f-{\bf\Delta};\
{\bf k}'={\bf q}_b+{\bf\Delta};\ {\bf\Delta}={\bf
p}_F-{\bf p}_{B}$;
$$\Lambda^{(-)}(p)=\frac{\epsilon(p)-\bigl( m\gamma
^0+\gamma^0({\bm{ \gamma}{\bf p}})\bigr)}{ 2\epsilon (p)}.$$
Here \cite{f} 
\begin{eqnarray*} 
p_{f,q}&=&\epsilon_{f,q}(p)\frac{p_F}{M_F}
\pm\sum_{i=1}^3 n^{(i)}(p_F)p^i,\\
q_{b,q}&=&\epsilon_{b,q}(q)\frac{p_{B}}{M_{B}} \pm \sum_{i=1}^3 n^{(i)}
(p_{B})q^i,\end{eqnarray*}
and $n^{(i)}$ are three four-vectors given by
$$ n^{(i)\mu}(p)=\left\{ \frac{p^i}{M},\ \delta_{ij}+
\frac{p^ip^j}{M(E+M)}\right\}, \quad E=\sqrt{{\bf p}^2+M^2}.$$

It is important to note that the wave functions entering the weak current
matrix element (\ref{mxet}) are not in the rest frame in general. For example, 
in the $B$ meson rest frame (${\bf p}_{B}=0$), the final  meson
is moving with the recoil momentum ${\bf \Delta}$. The wave function
of the moving  meson $\Psi_{F\,{\bf\Delta}}$ is connected 
with the  wave function in the rest frame 
$\Psi_{F\,{\bf 0}}\equiv \Psi_F$ by the transformation \cite{f}
\begin{equation}
\label{wig}
\Psi_{F\,{\bf\Delta}}({\bf
p})=D_f^{1/2}(R_{L_{\bf\Delta}}^W)D_q^{1/2}(R_{L_{
\bf\Delta}}^W)\Psi_{F\,{\bf 0}}({\bf p}),
\end{equation}
where $R^W$ is the Wigner rotation, $L_{\bf\Delta}$ is the Lorentz boost
from the meson rest frame to a moving one, and   
the rotation matrix $D^{1/2}(R)$ in spinor representation is given by
\begin{equation}\label{d12}
{1 \ \ \,0\choose 0 \ \ \,1}D^{1/2}_{q,f}(R^W_{L_{\bf\Delta}})=
S^{-1}({\bf p}_{q,f})S({\bf\Delta})S({\bf p}),
\end{equation}
where
$$
S({\bf p})=\sqrt{\frac{\epsilon(p)+m}{2m}}\left(1+\frac{\bm{\alpha}{\bf p}}
{\epsilon(p)+m}\right)
$$
is the usual Lorentz transformation matrix of the four-spinor.

\section{Form factors of rare semileptonic decays}
\label{ffr}

The matrix elements of the weak current for rare $B$ decays ($B$
denotes either $B$ or $B_c$) to pseudoscalar mesons ($P=K,D_s,D$) 
can be parametrized by three invariant form factors,
\begin{eqnarray}
  \label{eq:pff1}
  \langle P(p_F)|\bar q \gamma^\mu b|B(p_B)\rangle
  &=&f_+(q^2)\left[p_B^\mu+ p_F^\mu-\frac{M_B^2-M_P^2}{q^2}\ q^\mu\right]+
  f_0(q^2)\frac{M_B^2-M_P^2}{q^2}\ q^\mu,\qquad\\ \cr
\label{eq:pff2}
\langle P(p_F)|\bar q \sigma^{\mu\nu}q_\nu b|B(p_B)\rangle&=&
\frac{if_T(q^2)}{M_B+M_P} [q^2(p_B^\mu+p_F^\mu)-(M_B^2-M_P^2)q^\mu],
\end{eqnarray}
where $f_+(0)=f_0(0)$, $q=p_B- p_F$, and $M_{B,P}$ are the masses of the $B$ meson  
and pseudoscalar $P$ meson, respectively.

The corresponding matrix elements for the rare $B$ decays to vector
mesons ($V=K^*,D_s^*,D^*$) are parametrized by seven form factors,
\begin{eqnarray}
  \label{eq:vff1}
  \langle V(p_F)|\bar q \gamma^\mu b|B(p_B)\rangle&=
  &\frac{2iV(q^2)}{M_B+M_V} \epsilon^{\mu\nu\rho\sigma}\epsilon^*_\nu
  p_{B\rho} p_{F\sigma},\\ \cr
\label{eq:vff2}
\langle V(p_F)|\bar q \gamma^\mu\gamma_5 b|B(p_B)\rangle&=&2M_V
A_0(q^2)\frac{\epsilon^*\cdot q}{q^2}\ q^\mu
 +(M_B+M_V)A_1(q^2)\left(\epsilon^{*\mu}-\frac{\epsilon^*\cdot
    q}{q^2}\ q^\mu\right)\cr\cr
&&-A_2(q^2)\frac{\epsilon^*\cdot q}{M_B+M_V}\left[p_B^\mu+
  p_F^\mu-\frac{M_B^2-M_V^2}{q^2}\ q^\mu\right], \\\cr
\label{eq:vff3}
\langle V(p_F)|\bar q i\sigma^{\mu\nu}q_\nu b|B(p_B)\rangle&=&2T_1(q^2)
\epsilon^{\mu\nu\rho\sigma} \epsilon^*_\nu p_{F\rho} p_{B\sigma},\\\cr
\label{eq:vff4}
\langle V(p_F)|\bar q i\sigma^{\mu\nu}\gamma_5q_\nu b|B(p_B)\rangle&=&
T_2(q^2)[(M_B^2-M_V^2)\epsilon^{*\mu}-(\epsilon^*\cdot q)(p_B^\mu+
p_F^\mu)]\cr\cr
&&+T_3(q^2)(\epsilon^*\cdot q)\left[q^\mu-\frac{q^2}{M_B^2-M_V^2}
  (p_B^\mu+p_F^\mu)\right], 
\end{eqnarray}
where $2M_VA_0(0)=(M_B+M_V)A_1(0)-(M_B-M_V)A_2(0)$, $T_1(0)=T_2(0)$; 
$M_V$ and $\epsilon_\mu$ are the mass and polarization vector of
the final vector meson. 

We previously studied the form factors ($f_+,f_0,V,A_0,A_1,A_2$)
parametrizing the matrix elements of
vector and axial vector charged weak currents for $B\to \pi(\rho)$
\cite{bdecays} and $B_c\to \eta_c(J/\psi)$, $B_c\to
D^{(*)}$ \cite{bcjpsi}, $B_c\to B_s^{(*)}(B^{(*)})$ \cite{bcbs}
 transitions in the framework of our model. The necessary
formulas for these form 
factors can be found in Appendix of Ref.~\cite{bcjpsi}. Now we apply them
to the calculation of the  form factors, parametrizing neutral current
matrix elements for the $B\to K^{(*)}$ and $B_c\to D_s^{(*)}$,
$B_c\to D^{(*)}$ transitions. For the remaining tensor form factors we
use the same approach described in detail in
Refs.~\cite{bcjpsi,bcbs,bdecays}. Namely, we calculate exactly the
contribution of the leading vertex function $\Gamma^{(1)}$ 
(\ref{gamma1}) to the transition matrix element of the weak
current (\ref{mxet}) using the $\delta$-function.  For the evaluation of
the subleading contribution $\Gamma^{(2)}$  we use expansions in inverse powers of the
heavy $b$-quark mass from  the initial $B$ meson and of the large recoil
energy of the final heavy-light meson. Note that the latter
contributions turn out to be rather small numerically. Therefore we
obtain reliable expressions for the form factors in the whole
accessible kinimatical range. It is important to emphasize that doing
these calculations we consistently take into account all relativistic
corrections including boosts of the meson wave functions from the rest
frame to the moving one, given
by Eq.~(\ref{wig}).   The obtained expressions for the tensor
form factors $f_T,T_1,T_2,T_3$ are presented in Appendix (to simplify
these expressions the long-range anomalous chromomagnetic quark moment
was explicitly set as $\kappa=-1$). In
the limits of the infinitely heavy quark mass and large energy of the
final meson, the form
factors in our model satisfy all model independent symmetry relations
\cite{clopr,ffhm}.  

 For numerical calculations of the form factors we use the
quasipotential wave functions of the 
$B$, $B_c$, $K^*$, $D_s$ and $D$ mesons obtained in their mass spectra
calculations \cite{mass,hlm}. Our results for the masses of these
mesons are in good agreement with experimental data \cite{pdg}, which
we use in our calculations.

We find that the rare semileptonic $B\to K^{(*)} l^+l^-$ and $B_c\to D_s^{(*)}(
D^{(*)}) l^+l^-$ decay form factors can 
be approximated with good accuracy by the following expressions
\cite{melstech,bdecays}: 

(a) $F(q^2)= \{f_+(q^2),f_T(q^2),V(q^2),A_0(q^2),T_1(q^2)\}$ 
\begin{equation}
  \label{fitfv}
  F(q^2)=\frac{F(0)}{\displaystyle\left(1-\frac{q^2}{\tilde M^2}\right)
    \left(1-\sigma_1 
      \frac{q^2}{M_{B_s^*}^2}+ \sigma_2\frac{q^4}{M_{B_s^*}^4}\right)},
\end{equation}

\begin{figure}
  \centering
  \includegraphics[width=7.5cm]{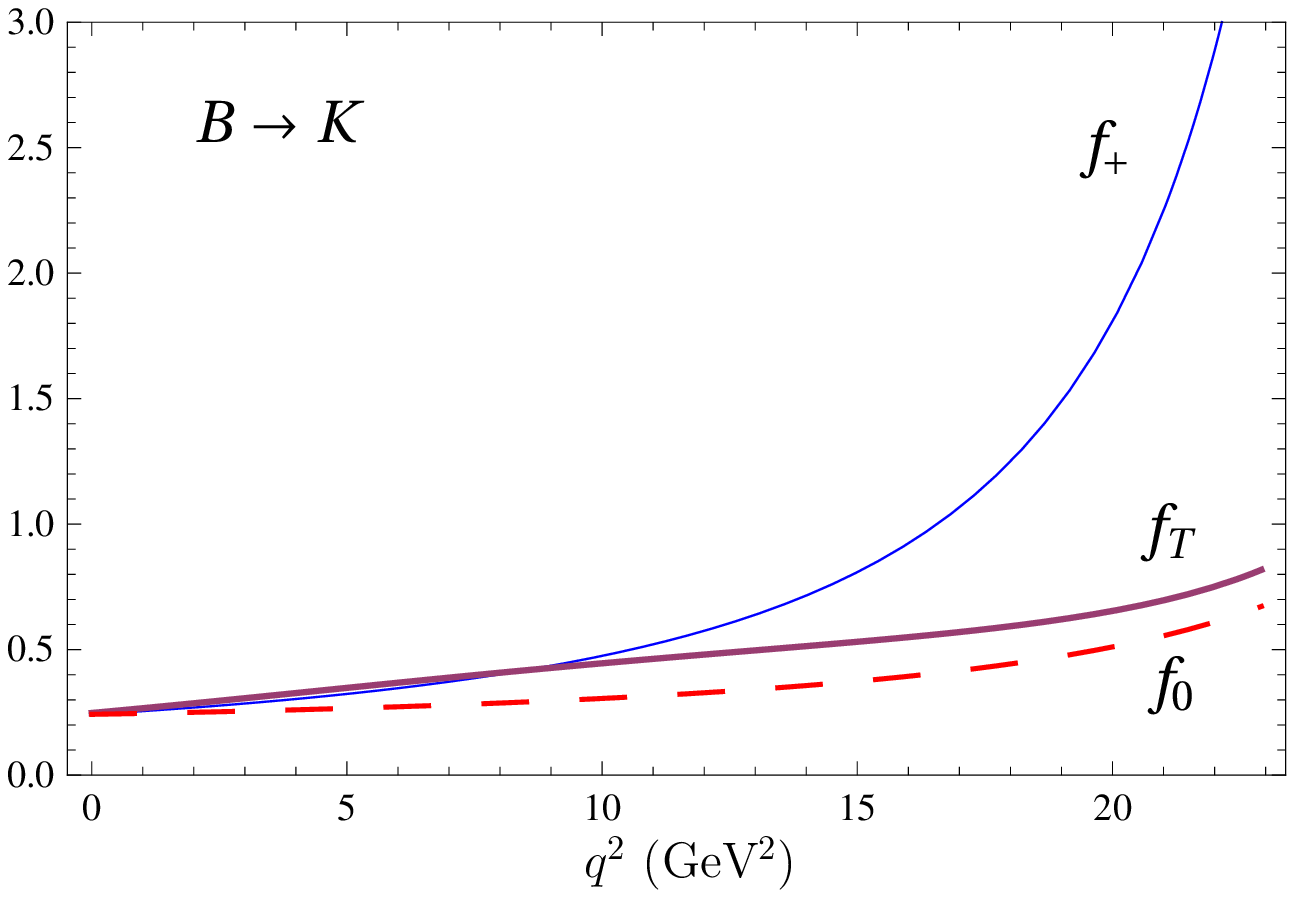} 

\vspace*{0.5cm}

\includegraphics[width=7.5cm]{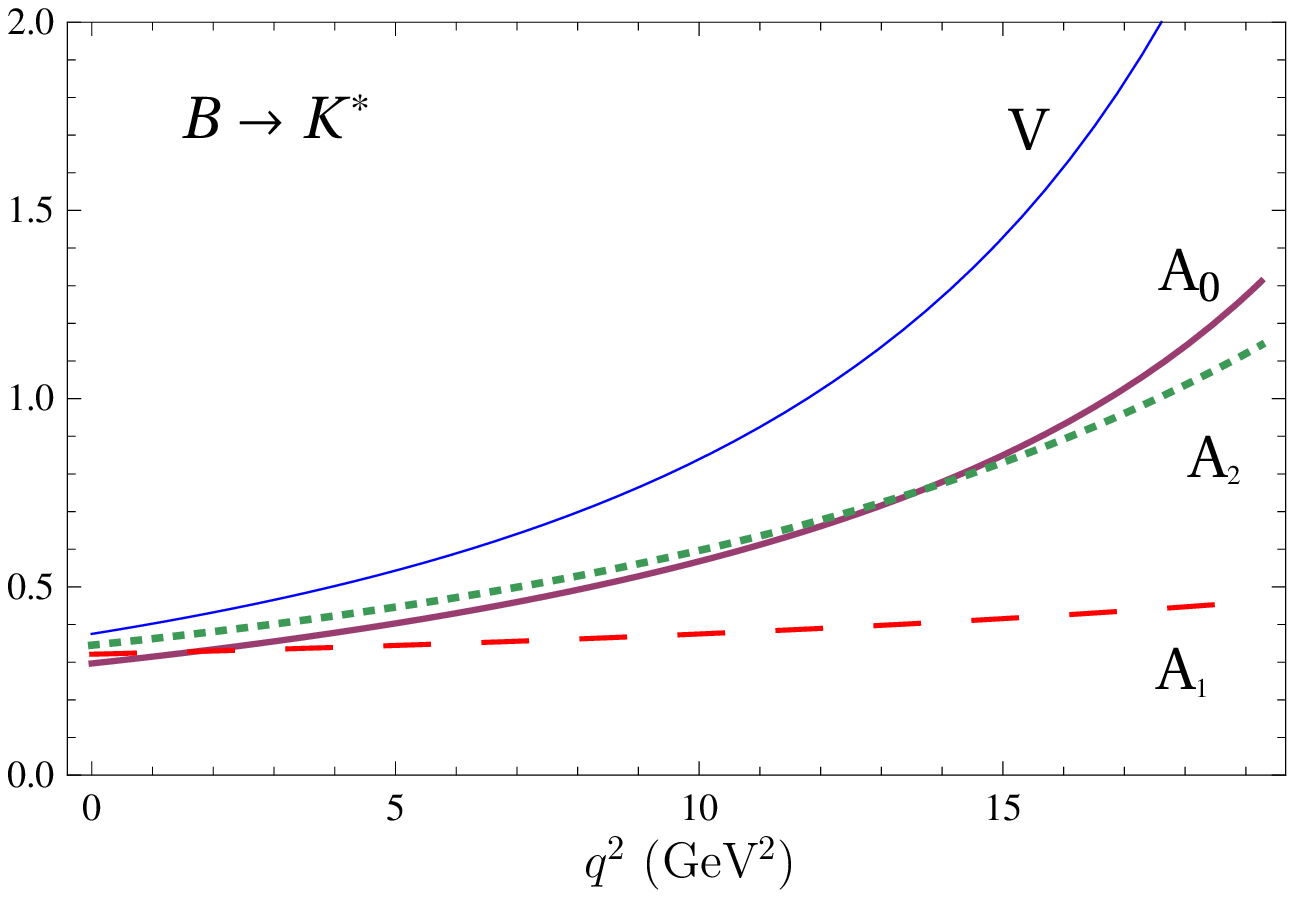}\ \ \ \
\  \includegraphics[width=7.5cm]{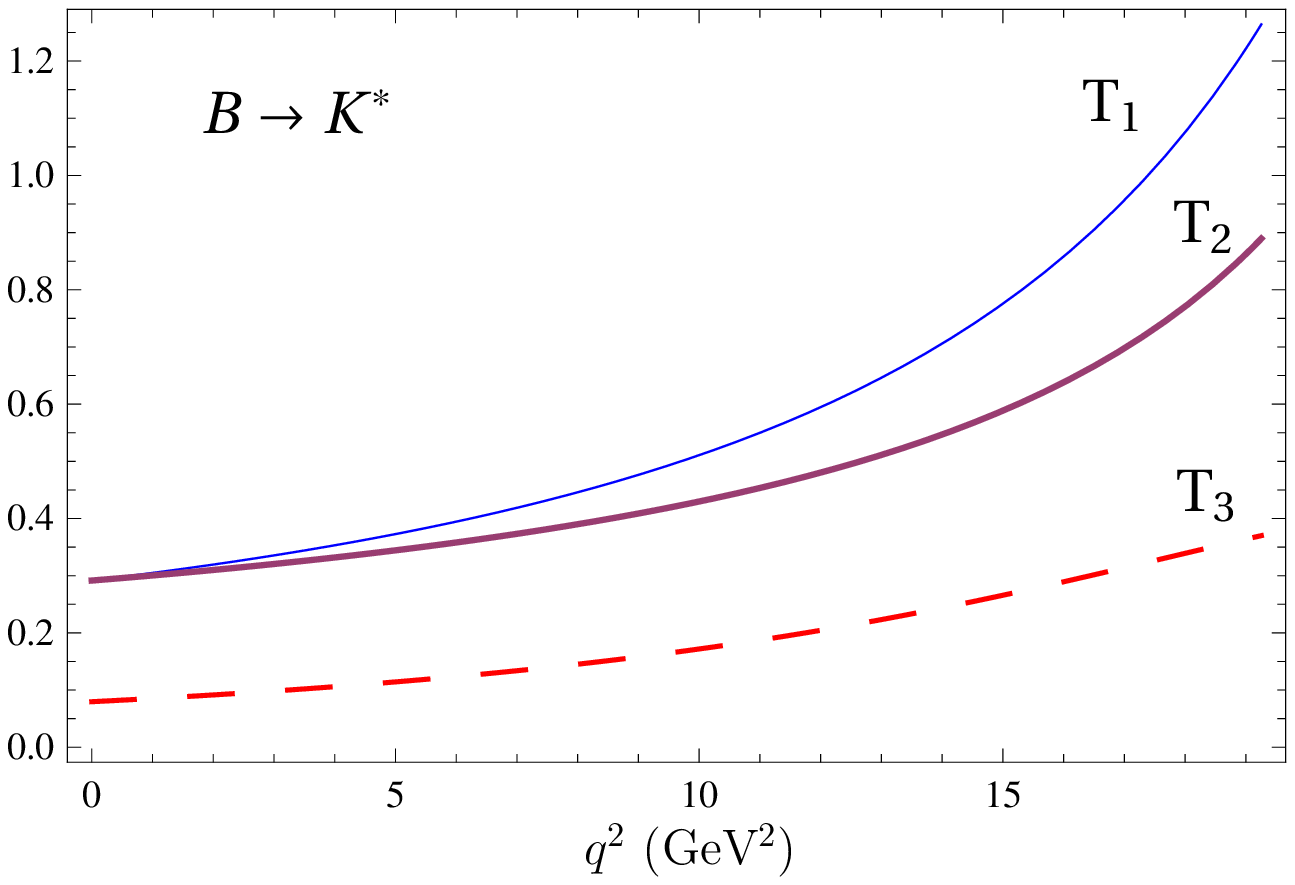}
  \caption{Form factors of the $B \to K^{(*)}$ decays.}
  \label{fig:ffbk}
\end{figure}

\begin{figure}
  \centering
  \includegraphics[width=7.5cm]{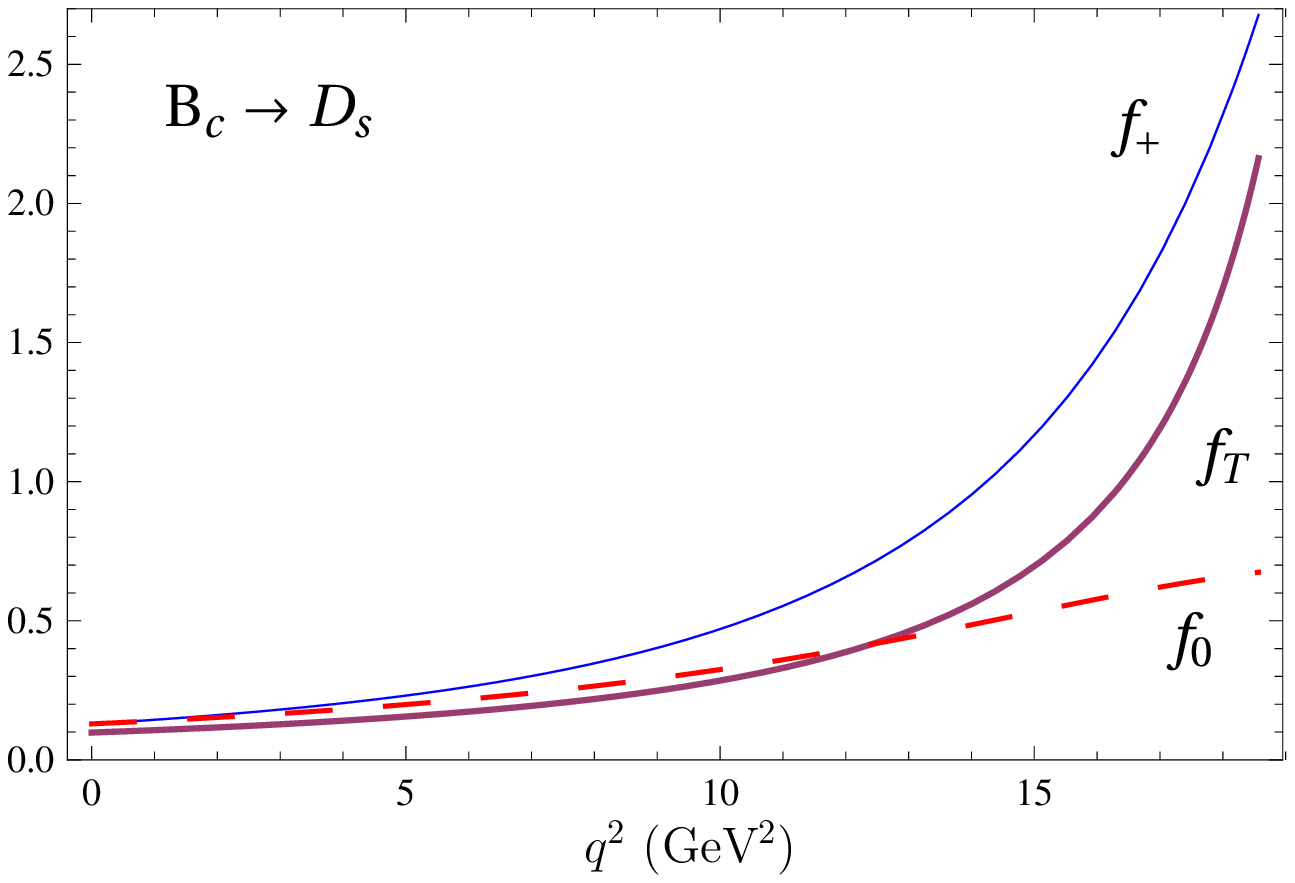} 

\vspace*{0.5cm}

\includegraphics[width=7.5cm]{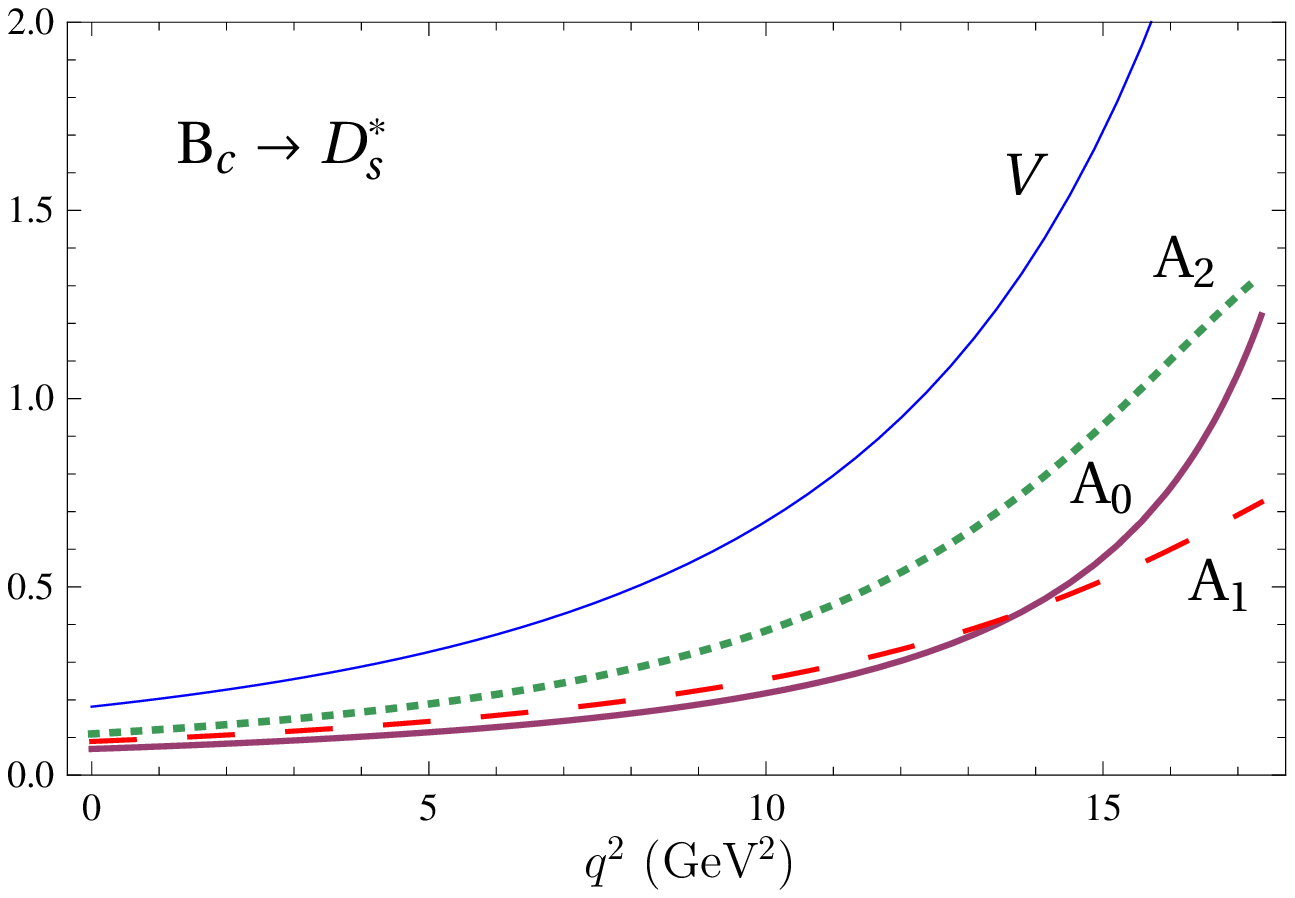}\ \ \ \
\  
\includegraphics[width=7.5cm]{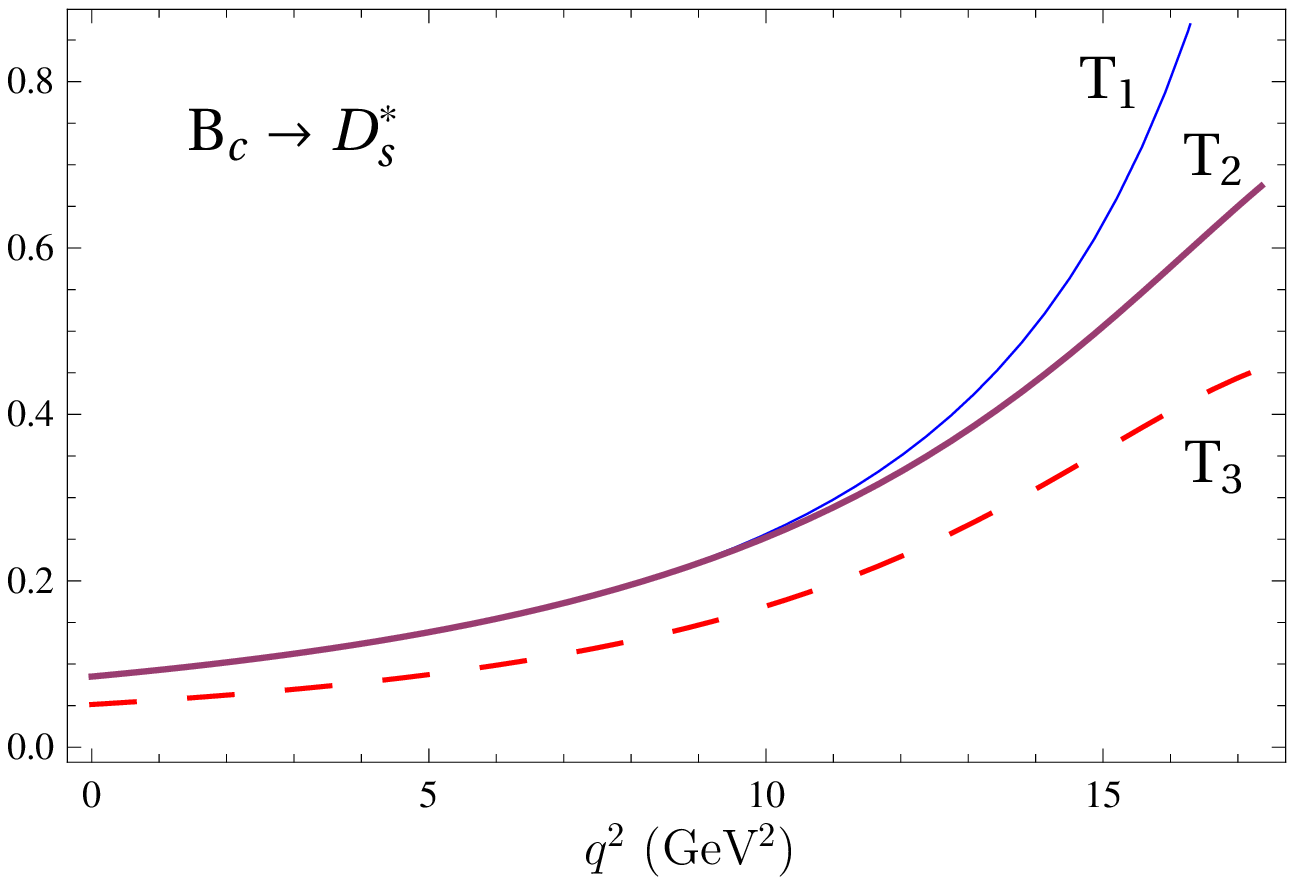}
  \caption{Form factors of the $B_c \to D_s^{(*)}$ decays.}
  \label{fig:ffbc}
\end{figure}

(b) $F(q^2)=\{f_0(q^2), A_1(q^2),A_2(q^2),T_2(q^2),T_3(q^2)\}$
\begin{equation}
  \label{fita12}
  F(q^2)=\frac{F(0)}{\displaystyle \left(1-\sigma_1
      \frac{q^2}{M_{B_s^*}^2}+ \sigma_2\frac{q^4}{M_{B_s^*}^4}\right)},
\end{equation}
where $\tilde M=M_{B_s}$ for $A_0$ and $\tilde M=M_{B_s^*}$ for all other
form factors (for $B_c\to D^{(*)} l^+l^-$ decays $M_{B^{(*)}_s}$
should be replaced by $M_{B^{(*)}}$).  The values  $F(0)$ and $\sigma_{1,2}$ are given in 
Tables~\ref{bkff}-\ref{bcdff}. The difference of fitted form factors from the
calculated ones does not exceed  1\%. We plot these form factors in
Figs.~\ref{fig:ffbk} and \ref{fig:ffbc}.

\begin{table}
\caption{Form factors of the rare semileptonic decays $B\to K^{(*)} l^+l^-$ 
  calculated in our model. Form factors $f_+(q^2)$, $f_T(q^2)$, $V(q^2)$,
  $A_0(q^2)$, $T_1(q^2)$ are fitted by Eq.~(\ref{fitfv}), and form factors
 $f_0(q^2)$, $A_1(q^2)$, $A_2(q^2)$, $T_2(q^2)$, $T_3(q^2)$ are fitted by Eq.~(\ref{fita12}).  }
\label{bkff}
\begin{ruledtabular}
\begin{tabular}{ccccccccccc}
   &\multicolumn{3}{c}{{$B\to K$}}&\multicolumn{7}{c}{{\  $B\to K^*$
     }}\\
\cline{2-4} \cline{5-11}
& $f_+$ & $f_0$& $f_T$& $V$ & $A_0$ &$A_1$&$A_2$&$T_1$&$T_2$&$T_3$ \\
\hline
$F(0)$&0.242&0.242&0.258 &  0.375 & 0.297& 0.321 & 0.345&0.291&0.291&0.080\\
$\sigma_1$&0.480&0.445& 1.198 &  1.019& 0.695&0.374&1.422&0.275&0.855&1.982\\
$\sigma_2$ &-0.537& -0.476&2.168&0.229&0.322&-0.138&0.548&-0.339&-0.256&1.198\\
\end{tabular}
\end{ruledtabular}
\end{table}

 \begin{table}
\caption{Form factors of the rare semileptonic decays $B_c\to D_s^{(*)} l^+l^-$ 
  calculated in our model. Form factors $f_+(q^2)$, $f_T(q^2)$, $V(q^2)$,
  $A_0(q^2)$, $T_1(q^2)$ are fitted by Eq.~(\ref{fitfv}), and form factors
 $f_0(q^2)$, $A_1(q^2)$, $A_2(q^2)$, $T_2(q^2)$, $T_3(q^2)$ are fitted by Eq.~(\ref{fita12}).  }
\label{bcdsff}
\begin{ruledtabular}
\begin{tabular}{ccccccccccc}
   &\multicolumn{3}{c}{{$B_c\to D_s$}}&\multicolumn{7}{c}{{\  $B_c\to D_s^*$
     }}\\
\cline{2-4} \cline{5-11}
& $f_+$ & $f_0$& $f_T$& $V$ & $A_0$ &$A_1$&$A_2$&$T_1$&$T_2$&$T_3$ \\
\hline
$F(0)$&0.129&0.129&0.098 &  0.182 & 0.070& 0.089 & 0.110&0.085&0.085&0.051\\
$\sigma_1$&2.096&2.331& 1.412 &  2.133& 1.561&2.479&2.833&1.540&2.577&2.783\\
$\sigma_2$ &1.147& 1.666&0.048&1.183&0.192&1.686&2.167&0.248&1.859&2.170\\
\end{tabular}
\end{ruledtabular}
\end{table}

\begin{table}
\caption{Form factors of the rare semileptonic decays $B_c\to D^{(*)} l^+l^-$ 
  calculated in our model. Form factors $f_+(q^2)$, $f_T(q^2)$, $V(q^2)$,
  $A_0(q^2)$, $T_1(q^2)$ are fitted by Eq.~(\ref{fitfv}), and form factors
 $f_0(q^2)$, $A_1(q^2)$, $A_2(q^2)$, $T_2(q^2)$, $T_3(q^2)$ are fitted by Eq.~(\ref{fita12}).  }
\label{bcdff}
\begin{ruledtabular}
\begin{tabular}{ccccccccccc}
   &\multicolumn{3}{c}{{$B_c\to D$}}&\multicolumn{7}{c}{{\  $B_c\to D^*$
     }}\\
\cline{2-4} \cline{5-11}
& $f_+$ & $f_0$& $f_T$& $V$ & $A_0$ &$A_1$&$A_2$&$T_1$&$T_2$&$T_3$ \\
\hline
$F(0)$&0.081&0.081&0.061 &  0.125 & 0.035& 0.054 & 0.071&0.055&0.055&0.034\\
$\sigma_1$&2.167&2.455& 1.363 &  2.247& 1.511&2.595&2.800&1.520&2.633&2.801\\
$\sigma_2$ &1.203& 1.729&0.026&1.346&0.175&1.784&2.073&0.207&1.886&2.108\\
\end{tabular}
\end{ruledtabular}
\end{table}

\section{Results and discussion}
 \label{sec:rd}

Now we use the obtained form factors for the numerical calculation of decay
rates and other important observables of the rare semileptonic $B$
decays and confront their values with available experimental data. 

\subsection{$B\to P l^+l^-$ and $B\to V l^+l^-$ decays} 

The matrix element of the $b\to f l^+l^-$ ($f=s$ or $d$) decay amplitude (\ref{eq:wda}) between meson states can be written \cite{fgikl,abhh} in
the following form
\begin{eqnarray}
  \label{eq:mtl}
  {\cal M}(B\to Pl^+l^-)&=&\frac{G_F\alpha}{2\sqrt{2}\pi}
  |V_{tf}^*V_{tb}|\left[T^{(1)}_\mu(\bar l\gamma^\mu l)+T^{(2)}_\mu
    (\bar l\gamma^\mu \gamma_5l)\right],\cr
 {\cal M}(B\to Vl^+l^-)&=&\frac{G_F\alpha}{2\sqrt{2}\pi}
  |V_{tf}^*V_{tb}|\left[\epsilon^{\dag\nu}T^{(1)}_{\mu\nu}(\bar l\gamma^\mu l)+\epsilon^{\dag\nu}T^{(2)}_{\mu\nu}
    (\bar l\gamma^\mu \gamma_5l)\right],
\end{eqnarray}
where $T^{(i)}$ are expressed through the form factors and the Wilson
coefficients. Then these amplitudes can be written in the helicity
basis $\varepsilon^{\mu}(m)$ as (see \cite{fgikl})

(a) $B\to P$ transition:
\begin{equation}
  \label{eq:hap}
  H^{(i)}_m=\varepsilon^{\dag \mu}(m)T^{(i)}_\mu,
\end{equation}
where  
\begin{eqnarray}
  \label{eq:hp}
  H^{(i)}_\pm&=&0,\cr
H^{(1)}_0&=&\frac{\lambda^{1/2}}{\sqrt{q^2}}\left[C_9^{eff} f_+(q^2)+ C_7^{eff}\frac{2m_b}{M_B+M_P}f_T(q^2)\right],\cr
H^{(2)}_0&=&\frac{\lambda^{1/2}}{\sqrt{q^2}} C_{10} f_+(q^2),\cr
H^{(1)}_t&=&\frac{M_B^2-M_P^2}{\sqrt{q^2}}C_9^{eff} f_0(q^2),\cr
H^{(2)}_t&=&\frac{M_B^2-M_P^2}{\sqrt{q^2}}C_{10} f_0(q^2).
\end{eqnarray}
Here $\lambda\equiv
\lambda(M_B^2,M_P^2,q^2)=M_B^4+M_P^4+q^4-2(M_B^2M_P^2+M_P^2q^2+M_B^2q^2)$
and the subscripts $\pm,0,t$ denote transverse, longitudinal and time helicity
components, respectively.

(b) $B\to V$ transition:
\begin{equation}
  \label{eq:hav}
  H^{(i)}_m=\varepsilon^{\dag \mu}(m)\epsilon^{\dag\nu}T^{(i)}_{\mu\nu},
\end{equation}
where  $\epsilon^{\nu}$ is the polarization vector of the vector $V$ meson and
\begin{eqnarray}
  \label{eq:hpv}
  H^{(1)}_\pm&=&-(M_B^2-M_V^2)\left[C_9^{eff}\frac{A_1(q^2)}{M_B-M_V}+
    \frac{2m_b}{q^2}C_7^{eff}T_2(q^2)\right]\cr
&&\pm \lambda^{1/2}
\left[C_9^{eff}\frac{V(q^2)}{M_B+M_V}+
    \frac{2m_b}{q^2}C_7^{eff}T_1(q^2)\right],\cr
 H^{(2)}_\pm&=&C_{10}\left[-(M_B+M_V)A_1(q^2)\pm \lambda^{1/2}
\frac{V(q^2)}{M_B+M_V}\right],\cr
H^{(1)}_0&=&-\frac{1}{2M_V\sqrt{q^2}}\Biggl[C_9^{eff}\left\{(M_B^2-M_V^2-q^2)
    (M_B+M_V)A_1(q^2)-\frac{\lambda}{M_B+M_V}A_2(q^2)\right\}\cr
&&+2m_b C_7^{eff}\left\{(M_B^2+3M_V^2-q^2)
 T_2(q^2)-\frac{\lambda}{M_B^2-M_V^2}T_3(q^2)\right\}\Biggr],\cr
H^{(2)}_0&=&-\frac{1}{2M_V\sqrt{q^2}}C_{10}\left[(M_B^2-M_V^2-q^2)
    (M_B+M_V)A_1(q^2)-\frac{\lambda}{M_B+M_V}A_2(q^2)\right],\cr
H^{(1)}_t&=&-\frac{\lambda^{1/2}}{\sqrt{q^2}}C_9^{eff} A_0(q^2),\cr
H^{(2)}_t&=&-\frac{\lambda^{1/2}}{\sqrt{q^2}}C_{10} A_0(q^2),
\end{eqnarray}
here $\lambda\equiv \lambda(M_B^2,M_V^2,q^2)=M_B^4+M_V^4+q^4-2(M_B^2M_V^2+M_V^2q^2+M_B^2q^2)$.

The differential decay rate then reads \cite{fgikl}
\begin{eqnarray}
  \label{eq:dgamma}
  \frac{d\Gamma(B\to P(V)l^+l^-)}{dq^2}&=&\frac{G_F^2}{(2\pi)^3}
  \left(\frac{\alpha |V_{tf}^*V_{tb}|}{2\pi}\right)^2
  \frac{\lambda^{1/2}q^2}{48M_B^3} \sqrt{1-\frac{4m_l^2}{q^2}}
  \Biggl[H^{(1)}H^{\dag(1)}\left(1+\frac{4m_l^2}{q^2}\right)\cr
&& +
    H^{(2)}H^{\dag(2)}\left(1-\frac{4m_l^2}{q^2}\right) +\frac{2m_l^2}{q^2}3 H^{(2)}_tH^{\dag(2)}_t\Biggr],
\end{eqnarray}
where $m_l$ is the lepton mass and
\begin{equation}
  \label{eq:hh}
  H^{(i)}H^{\dag(i)}\equiv H^{(i)}_+H^{\dag(i)}_++H^{(i)}_-H^{\dag(i)}_-+H^{(i)}_0H^{\dag(i)}_0.
\end{equation}

The forward-backward asymmetry is given by
\begin{equation}
  \label{eq:afb}
  A_{FB}=\frac34 \sqrt{1-\frac{4m_l^2}{q^2}}\frac{ {\rm Re}(H^{(1)}_+H^{\dag(2)}_+)-{\rm Re}(H^{(1)}_-H^{\dag(2)}_-)}{H^{(1)}H^{\dag(1)}\left(1+\frac{4m_l^2}{q^2}\right) +
    H^{(2)}H^{\dag(2)}\left(1-\frac{4m_l^2}{q^2}\right) +\frac{2m_l^2}{q^2}3 H^{(2)}_tH^{\dag(2)}_t}.
\end{equation}

The longitudinal fraction of the $V$ polarization has the form

\begin{equation}
  \label{eq:fl}
  F_L=\frac{H^{(1)}_0H^{\dag(1)}_0\left(1+\frac{4m_l^2}{q^2}\right) +
    H^{(2)}_0H^{\dag(2)}_0\left(1-\frac{4m_l^2}{q^2}\right) +\frac{2m_l^2}{q^2}3 H^{(2)}_tH^{\dag(2)}_t }{H^{(1)}H^{\dag(1)}\left(1+\frac{4m_l^2}{q^2}\right) +
    H^{(2)}H^{\dag(2)}\left(1-\frac{4m_l^2}{q^2}\right) +\frac{2m_l^2}{q^2}3 H^{(2)}_tH^{\dag(2)}_t}.
\end{equation}
These two observables are the most popular quantities for $B\to
K^*(\to K\pi)\mu^+\mu^-$ decays, since they are convenient for the experimental
measurements. They enter the decay differential distributions in $\cos
\theta_K$
\begin{equation}
  \label{eq:dgth}
  \frac1{\Gamma}\frac{d\Gamma(B\to K^*\mu^+\mu^-)}{d\cos \theta_K}=
  \frac32 F_L\cos^2 \theta_K+\frac34 (1- F_L)(1-\cos^2
  \theta_K), 
\end{equation}
and in $\cos \theta_\mu$
\begin{equation}
  \label{eq:dgtheta}
  \frac1{\Gamma}\frac{d\Gamma(B\to K^*\mu^+\mu^-)}{d\cos \theta_\mu}=
  \frac34 F_L(1-\cos^2 \theta_\mu)+\frac38 (1-F_L)(1+\cos^2
  \theta_\mu) +A_{FB}\cos \theta_\mu, 
\end{equation}
where $\theta_K$ is the angle between the kaon direction and the
direction opposite to the $B$ meson in the $K^*$ rest frame, and 
$\theta_\mu$ is the angle between the $\mu^+$ and the opposite of the
$B$ direction in the dilepton rest frame. Therefore they can be
determined experimentally using the angular analysis. 

\subsection{$B\to P \nu\bar\nu$ and $B\to V \nu\bar\nu$ decays} 

The differential decay rate for the $B\to P(V)\nu\bar\nu$ is given
by
\begin{equation}
  \label{eq:dgnunu}
  \frac{d\Gamma(B\to P(V)\nu\bar\nu)}{dq^2}=3\frac{G_F^2}{(2\pi)^3}
  \left(\frac{\alpha |V_{tf}^*V_{tb}|}{2\pi}\right)^2
  \frac{\lambda^{1/2}q^2}{24M_B^3}
  H^{(\nu)}H^{\dag(\nu)},
\end{equation}
where the factor of 3 originates from the sum over neutrino flavours,
\[  H^{(\nu)}H^{\dag(\nu)}\equiv
H^{(\nu)}_+H^{\dag(\nu)}_++H^{(\nu)}_-H^{\dag(\nu)}_-+H^{(\nu)}_0H^{\dag(\nu)}_0,\]
and the helicty amplitudes $H^{(\nu)}_m$ read as follows

(a) $B\to P$ transition: 
\begin{eqnarray}
  \label{eq:hpnu}
  H^{(\nu)}_\pm&=&0,\cr
H^{(\nu)}_0&=&\frac{\lambda^{1/2}}{\sqrt{q^2}} C_L^\nu f_+(q^2).
\end{eqnarray}

(b) $B\to V$ transition:
\begin{eqnarray}
  \label{eq:hpnuv}
 H^{(\nu)}_\pm&=&C_L^\nu\left[-(M_B+M_V)A_1(q^2)\pm \lambda^{1/2}
\frac{V(q^2)}{M_B+M_V}\right],\cr
H^{(\nu)}_0&=&-\frac{1}{2M_V\sqrt{q^2}}C_L^\nu\left[(M_B^2-M_V^2-q^2)
    (M_B+M_V)A_1(q^2)-\frac{\lambda}{M_B+M_V}A_2(q^2)\right].
\end{eqnarray}
Here $C_L^\nu=-X(x_t)/\sin^2\theta_W$, with $x_t=m_t^2/m_W^2$,
$\theta_W$ is the Weinberg angle, and the function $X(x_t)$ at the
leading-order in QCD has the form
\[X(x_t)=\frac{x}8\left(\frac{2+x}{x-1}+\frac{3x-6}{(x-1)^2}\ln x
\right),\]
while the next-to-leading order expressions are given in
Ref.~\cite{mu}.

Substituting the current experimental values for the top ($m_t$) and
$W$-boson ($m_W$) masses one gets \cite{absw}
\begin{equation}
  \label{eq:cl}
  C^\nu_L=-6.38\pm0.06,
\end{equation}
where the error is dominated by the top quark mass uncertainty. In the
following calculations we use the central value of $C^\nu_L$.

The differential  longitudinal polarization fraction $F_L$ of the $V$ meson is defined
similar to Eq.~(\ref{eq:fl})
\begin{equation}
  \label{eq:flnu}
  F_L=\frac{ H^{(\nu)}_0H^{\dag(\nu)}_0}{ H^{(\nu)}H^{\dag(\nu)}}.
\end{equation}

\subsection{Numerical results}

\begin{table}
\caption{Comparison of our predictions for the rare semileptonic $B\to
  K^{(*)}$ decay nonresonant branching fractions with experimental data (in $10^{-7}$). }
\label{brbk}
\begin{ruledtabular}
\begin{tabular}{lcccccc}
 Decay& our & BaBar \cite{babarbk,babarnu} &Belle \cite{bellebk,bellenu}& CDF
 \cite{cdfbk}& CDF \cite{cdfbk1} & HFAG \cite{hfag}\\
\hline
$B\to K \mu^+\mu^-$ & 4.19 &$3.4\pm0.7\pm0.2$ &
$4.8^{+0.5}_{-0.4}\pm0.3$ & $5.9\pm1.5\pm0.4$ & $3.8\pm0.5\pm0.3$ &
$4.5\pm0.4$\\
$B\to K \tau^+\tau^-$ & 1.17\\
$B\to K \nu\bar \nu$ & 26.1 & & $<140$\\
$B\to K^* \mu^+\mu^-$ & 9.25 &$7.8^{+1.9}_{-1.7}\pm1.1$ &
$10.7^{+1.1}_{-1.0}\pm0.9$ & $8.1\pm3.0\pm1.0$ & $10.6\pm1.4\pm0.9$ &
$10.8^{+1.2}_{-1.1}$\\
$B\to K^* \tau^+\tau^-$ & 1.03\\
$B\to K^* \nu\bar \nu$ & 63.2& $<800$\\
\end{tabular}
\end{ruledtabular}
\end{table}

Now we substitute the above calculated form factors
 in the expressions for decay rates, asymmetries and 
polarization fractions and perform numerical calculations. 

\begin{figure}
  \centering
 \includegraphics[width=7.5cm]{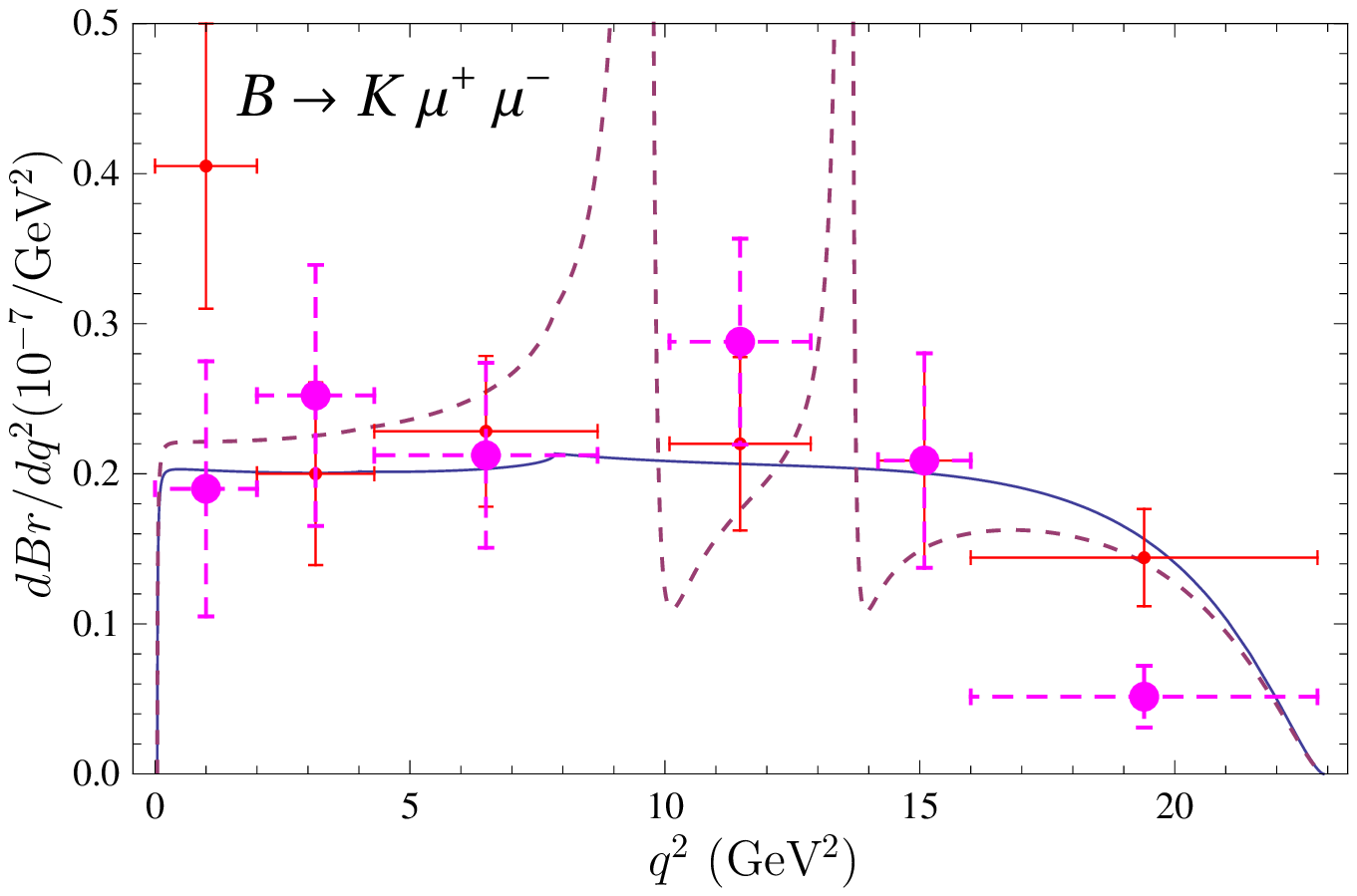} \ \ \  \ \ \includegraphics[width=7.5cm]{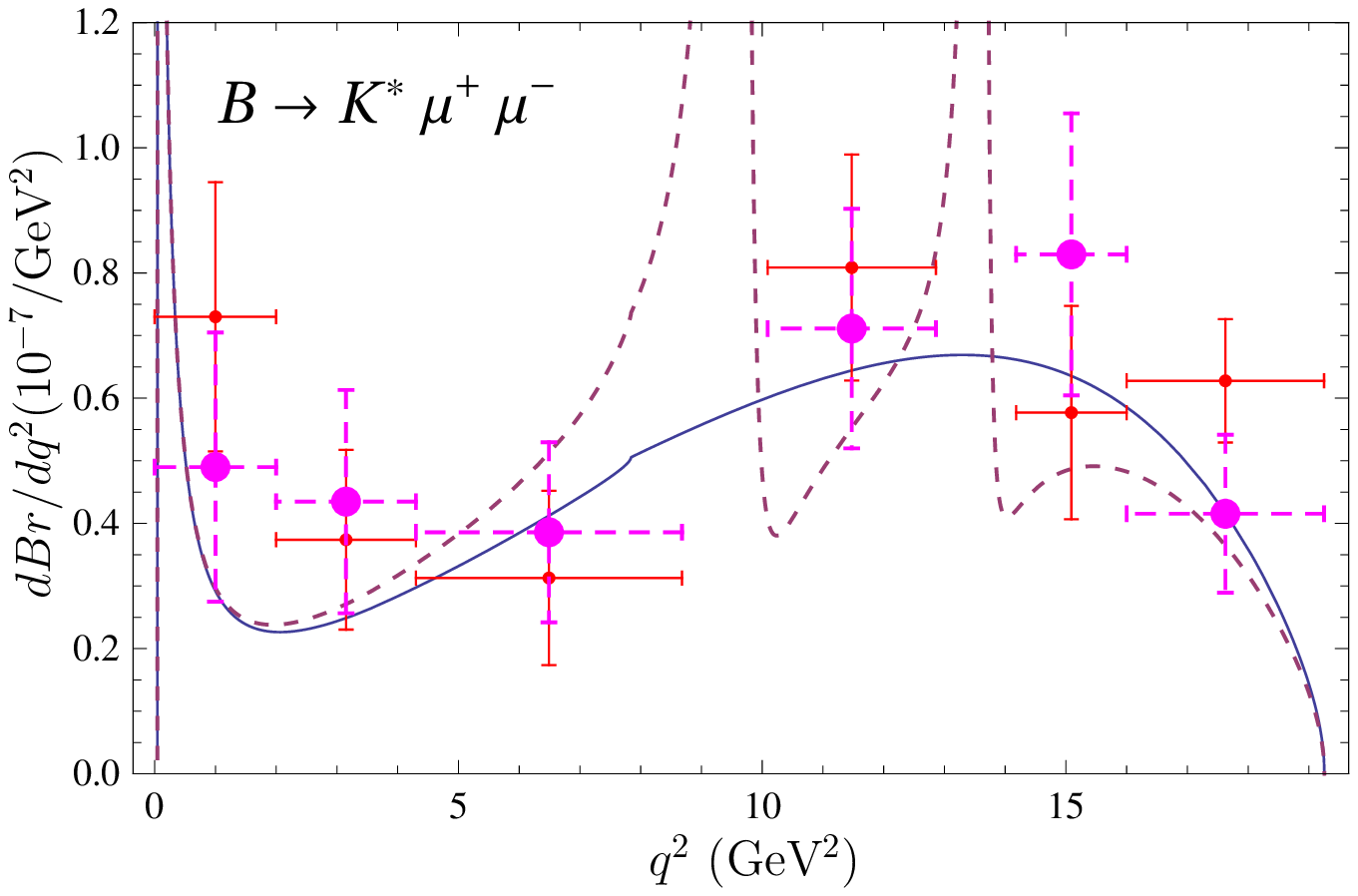}

\vspace*{0.5cm}

\includegraphics[width=7.5cm]{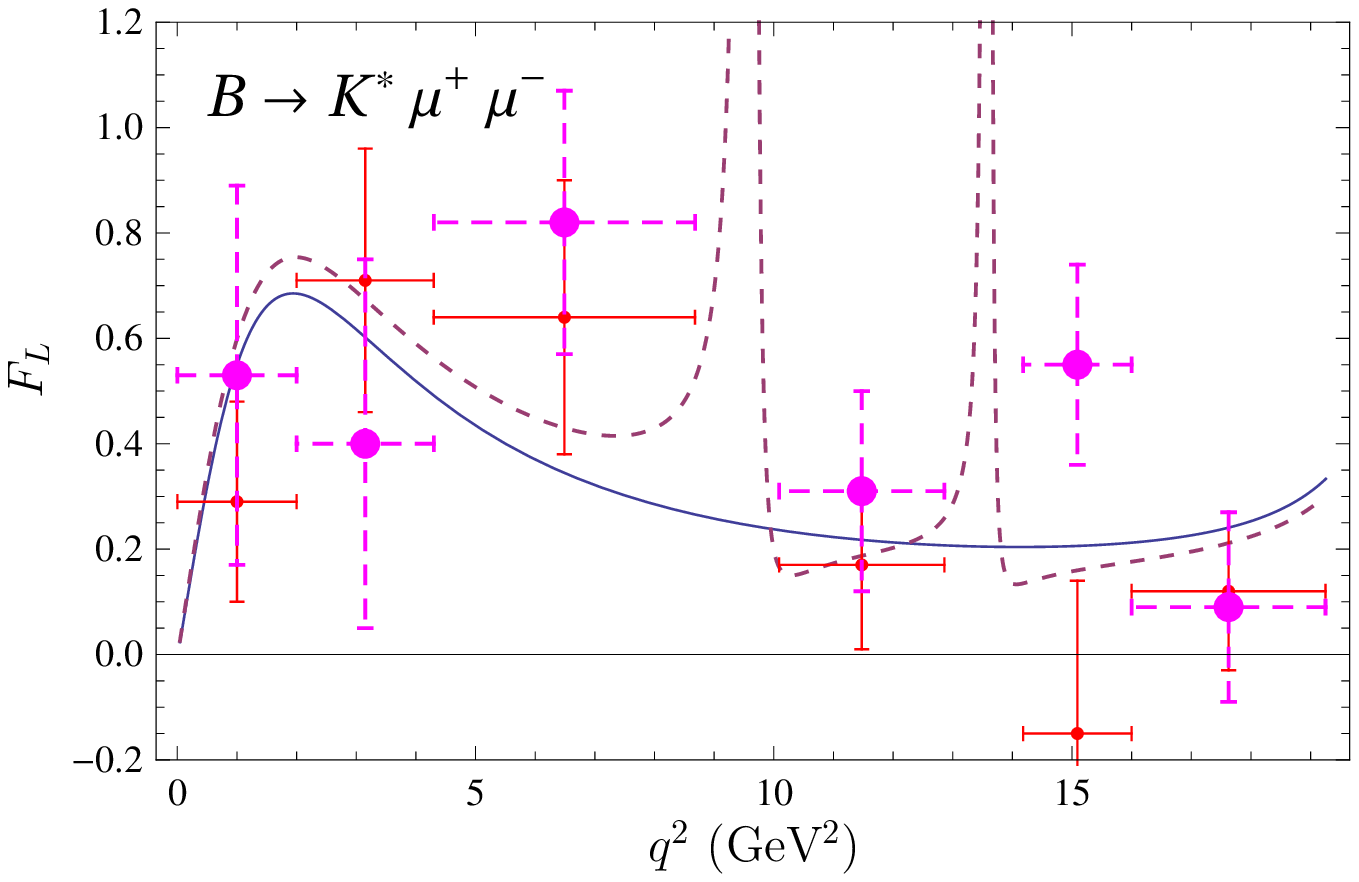}\ \ \ \
\  \includegraphics[width=7.5cm]{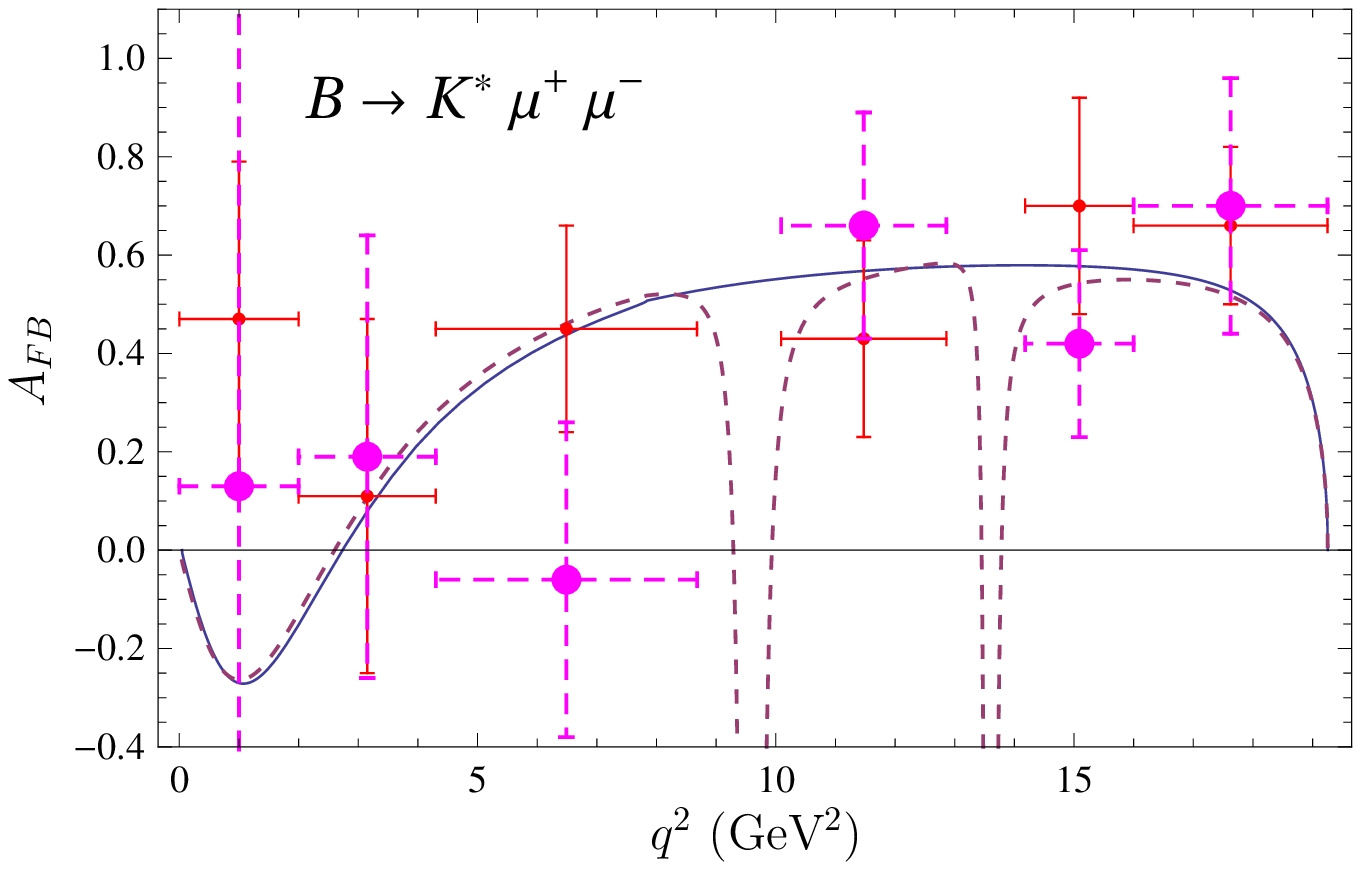}

  \caption{Comparison of theoretical predictions for the differential
    branching fractions $d Br/d q^2$, the $K^*$ longitudinal polarization
    $F_L$ and muon forward-backward asymmetry $A_{FB}$ for $B \to
    K^{(*)}$ decays with available experimental data. Nonresonant
    and resonant results are plotted by solid and dashed lines,
    respectively. Belle data are
    given by dots with solid error bars, while CDF data are presented by
    filled circles with dashed error bars.}
  \label{fig:brbk}
\end{figure}

First we compare the predictions of our model for the $B\to K l^+l^-$
and $B\to K^* l^+l^-$ decays with available 
experimental data. The calculated values for the branching
fractions of the  rare semileptonic decays $B\to K^{(*)} l^+l^-$ and $B\to K^{(*)} \nu\bar\nu$  and available experimental data are given in
Table~\ref{brbk}. We find good agreement of our results for 
 the $B\to K \mu^+\mu^-$ and $B\to
K^{*} \mu^+\mu^-$ decays with experimental data. A more stringent test of
our predictions can be achieved by comparison with new data on
differential decay distributions.  This is done in Fig.~\ref{fig:brbk}
where we confront our predictions
for differential decay rates, the  longitudinal polarization
fraction $F_L$ of the $K^*$ and the muon forward-backward asymmetry $A_{FB}$
with detailed experimental data from Belle \cite{bellebk} and CDF
\cite{cdfbk1}. In this figure we plot our results both for the nonresonant (solid
line) quantities and  quantities including the $J/\psi$ and $\psi'$ resonance contributions
(dashed line). Note that the resonance regions are vetoed in the
experimental analysis. Reasonable agreement of our predictions with
experimental data is found. The current experimental data on
$A_{FB}$ are not precise enough to give a definite conclusion whether this
asymmetry has a zero or not. Our model predicts  the
zero of $A_{FB}$ at  $q_0^2=2.74$~GeV$^2$ which is in agreement
with the value $q_0^2=2.88^{+0.44}_{-0.28}$~GeV$^2$ given in
\cite{abhh}. It is expected that the accuracy 
of experimental data will increase significantly in the near future.

\begin{figure}
  \centering
 \includegraphics[width=8cm]{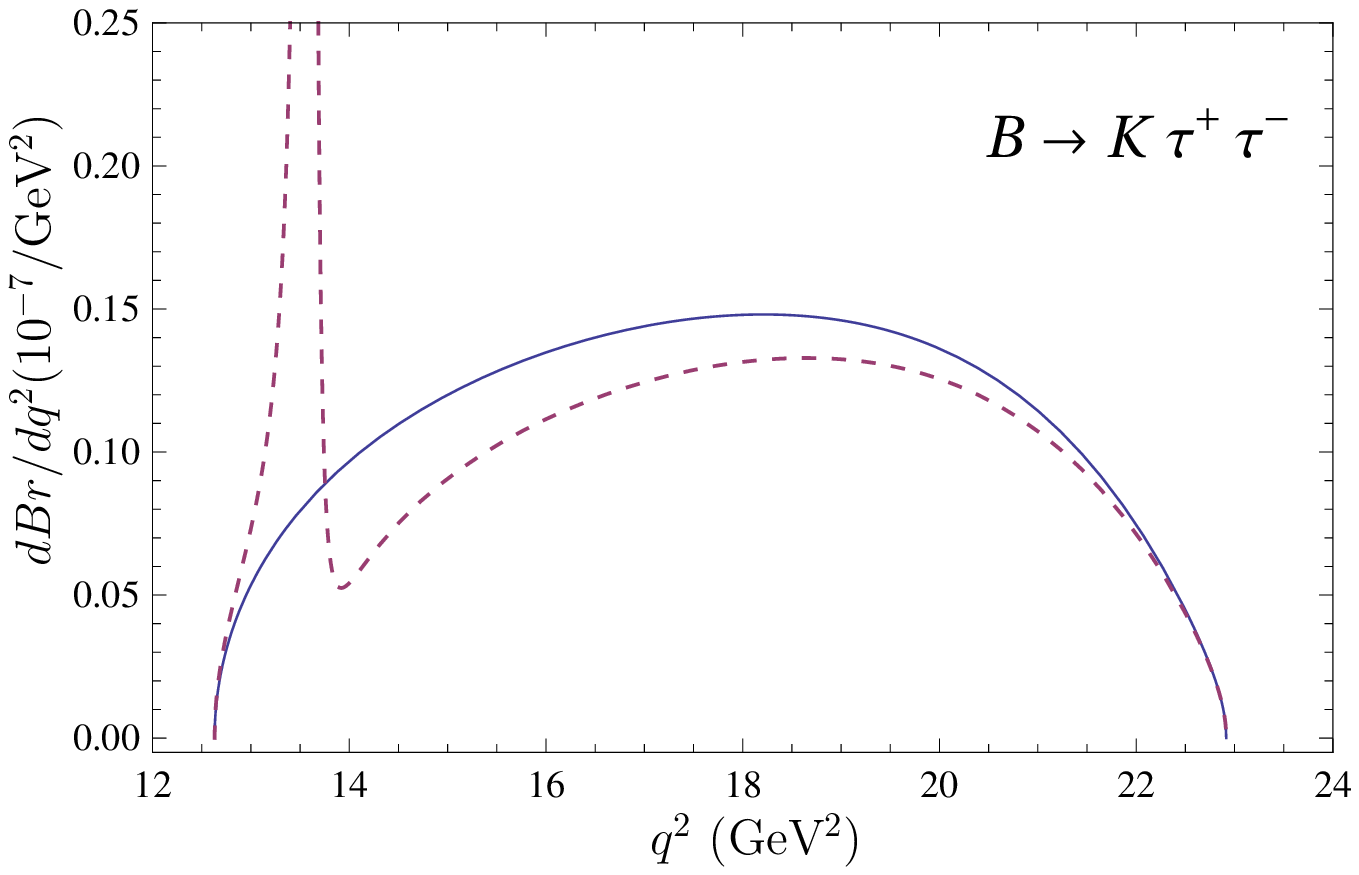}\ \ \  \includegraphics[width=8cm]{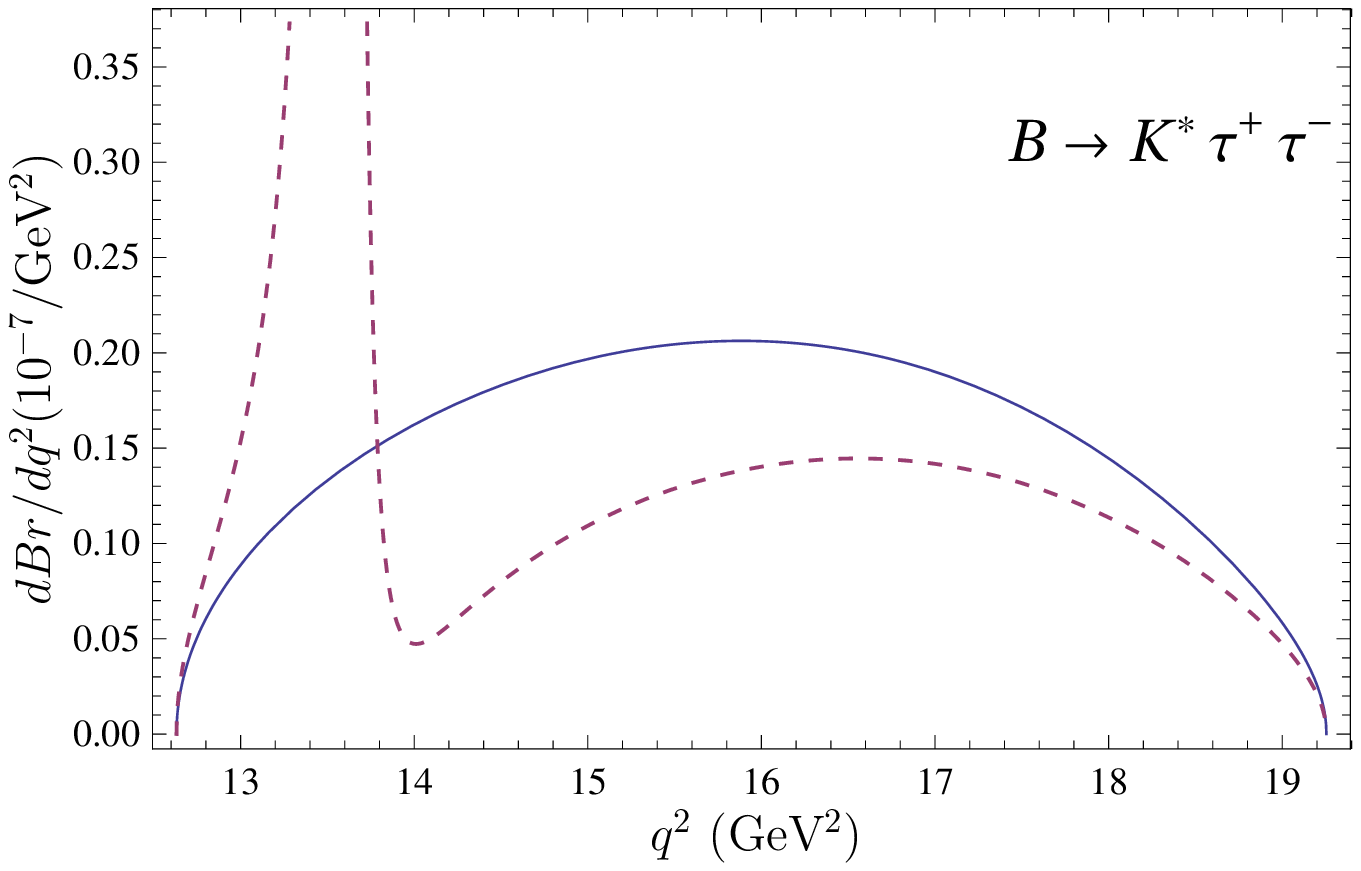}

  \caption{Predictions for the differential branching fractions  of $B \to
    K^{(*)}\tau^+\tau^-$ decays.  Nonresonant
    and resonant results are plotted by solid and dashed lines,
    respectively.}
  \label{fig:brbkt}
\end{figure}

\begin{figure}
  \centering
 \includegraphics[width=8cm]{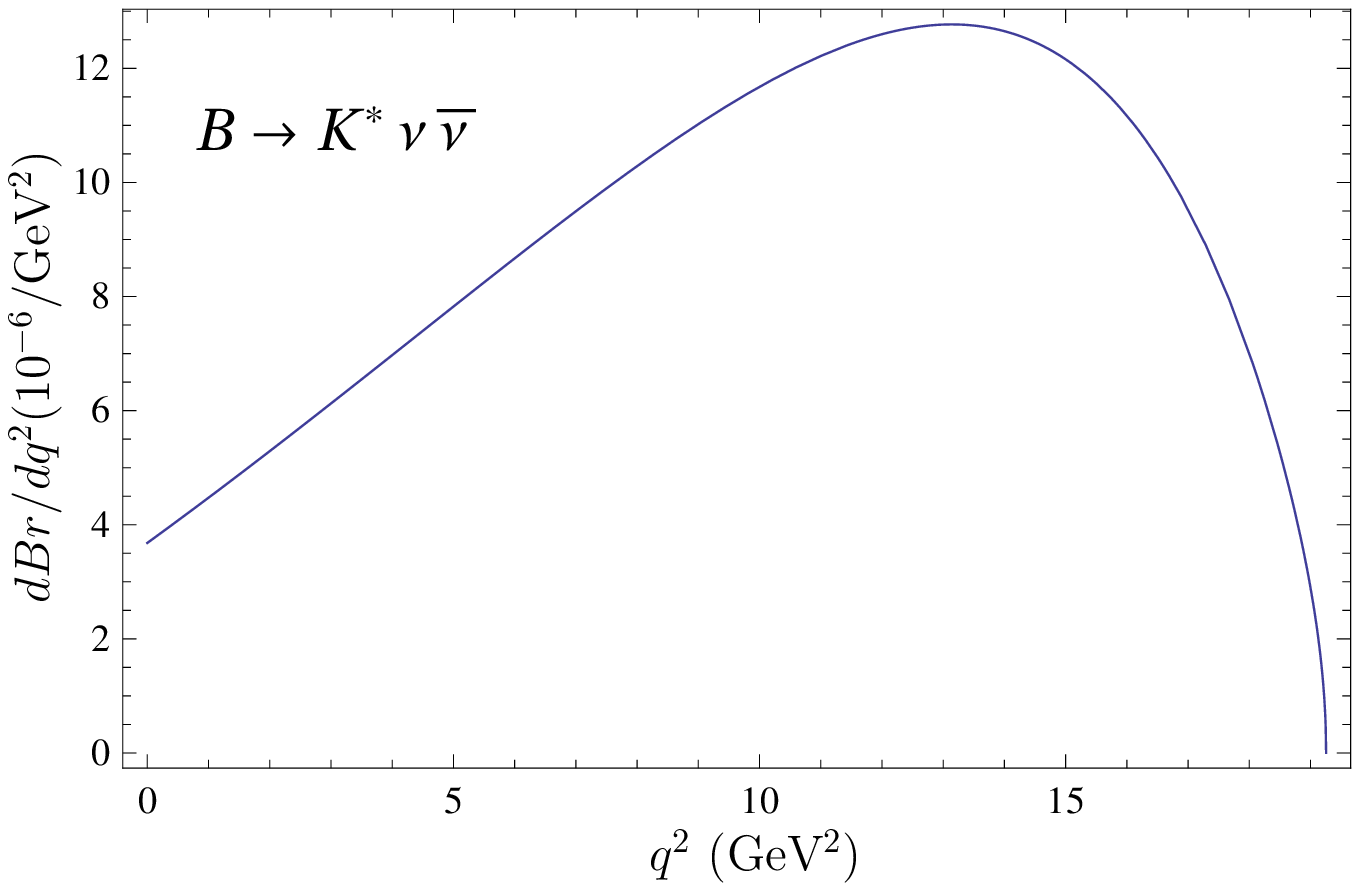}\ \ \  \includegraphics[width=8cm]{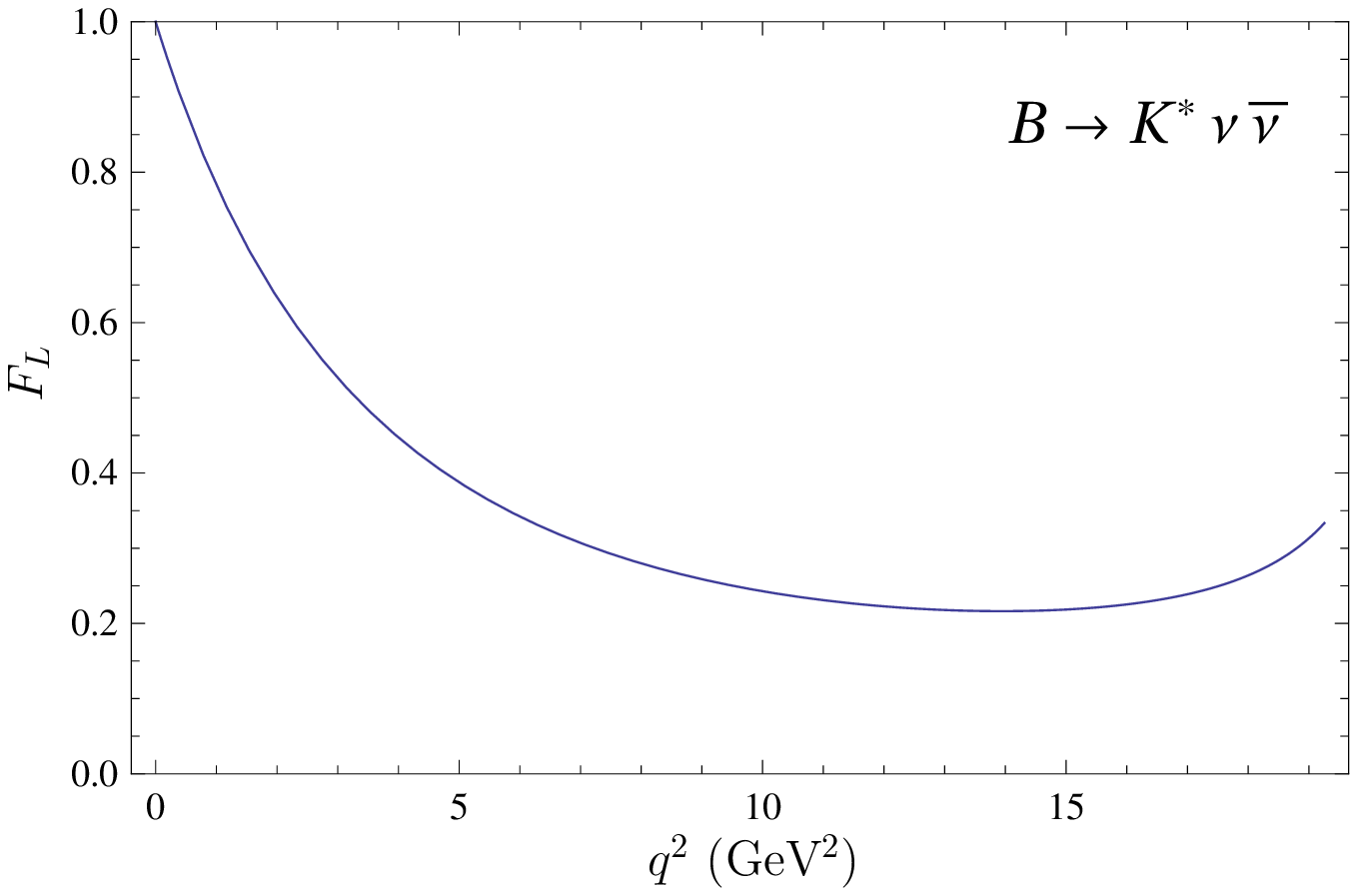}

  \caption{Predictions for the differential branching fractions and the
    $K^*$ longitudinal polarization fraction $F_L$  for the $B \to
    K^{*}\nu\bar\nu$ decay. }
  \label{fig:bknu}
\end{figure}

In Fig.~\ref{fig:brbkt} we plot our results for the differential
branching fractions of the $B \to K\tau^+\tau^-$ and $B \to K^{*}\tau^+\tau^-$ decays.
The calculated values for these decay branching fractions are presented in
Table~\ref{brbk}. There we also give our results for the $B\to K
\nu\bar\nu$ and $B\to K^* \nu\bar\nu$ branching fractions. None of these
modes have been measured yet. Only experimental upper bounds have been
recently set on branching fractions  for the $B\to K^*\nu\bar\nu$ decay by BaBar
\cite{babarnu} and for the $B\to K \nu\bar\nu$ decay by Belle
\cite{bellenu}. These bounds are about an order of magnitude higher
than our model predictions. In Fig.~\ref{fig:bknu} we show our
predictions for the differential branching fraction  and the $K^*$
longitudinal polarization fraction $F_L$ for the $B \to
K^{*}\nu\bar\nu$ decay. As it is noted in Ref.~\cite{absw}, the value
$F_L(0)=1$ is imposed by helicity conservation, while
$F_L(q^2_{\rm max})=1/3$ follows from the absence of a preferential
direction at the point $q^2_{\rm max}=(M_B-M_{K^*})^2$ where both $K^*$ and $B$ are at rest. The
differential branching fraction of the $B\to K^{*} \nu\bar\nu$ decay as
well as its integrated value are
close to the ones ($Br(B\to K^{*} \nu\bar\nu)=(6.8^{+1.0}_{-1.1})\times10^{-6}$) given in  Ref.~\cite{absw}, while the
shape of $F_L$ is slightly different. We also get the value for the $q^2$
integrated longitudinal polarization fraction
$\left<F_L\right>\approx0.31$, which is significantly lower than the one
of  Ref.~\cite{absw} $\left<F_L\right>=0.54\pm0.01$. Note that
our results for the branching fractions of the $ B\to K^{(*)}
\nu\bar\nu$ decay, though consistent with the ones from \cite{absw}, are
somewhat lower than the predictions from \cite{mns,bhi}.

\begin{table}
\caption{Comparison of theoretical predictions for the nonresonant branching
  fractions of the rare semileptonic $B_c\to
  D_s^{(*)}$ and $B_c\to D^{(*)}$ decays (in $10^{-8}$). }
\label{brbc}
\begin{ruledtabular}
\begin{tabular}{lccccc}
 Decay& our &  \cite{fgikl}& \cite{ghl}& 
 \cite{azizi}&  \cite{choi} \\
\hline
$B_c\to D_s \mu^+\mu^-$ & 11.6 &9.7 &13.6 & $6.1\pm1.5$ & 5.4\\
$B_c\to D_s \tau^+\tau^-$ & 3.3&2.2 & 3.4 & $2.3\pm0.5$& 1.4 \\
$B_c\to D_s \nu\bar \nu$ & 65& 73& 92 & $49\pm12$& 39\\
$B_c\to D_s^* \mu^+\mu^-$ & 21.2 &17.6 &40.9 & $29.9\pm5.0$ & \\
$B_c\to D_s^* \tau^+\tau^-$ & 3.5&2.2 & 5.1 & $2.05\pm0.76$&  \\
$B_c\to D_s^* \nu\bar \nu$ & 135& 142& 312 & & \\
$B_c\to D\mu^+\mu^-$ & 0.37 &0.44 &0.41 & $0.31\pm0.06$ & 0.18\\
$B_c\to D \tau^+\tau^-$ & 0.15&0.11 & 0.13 & $0.13\pm0.03$& 0.08 \\
$B_c\to D \nu\bar \nu$ & 2.16& 3.28& 2.77 & $3.48\pm0.71$& 1.31\\
$B_c\to D^* \mu^+\mu^-$ & 0.81 &0.71 &1.01 & $1.58\pm0.20$ & \\
$B_c\to D^* \tau^+\tau^-$ & 0.19&0.11 & 0.18 & $0.08-0.11$&  \\
$B_c\to D^* \nu\bar \nu$ & 5.12& 5.78& 7.64 & & \\
\end{tabular}
\end{ruledtabular}
\end{table}

\begin{figure}
  \centering
\includegraphics[width=7.5cm]{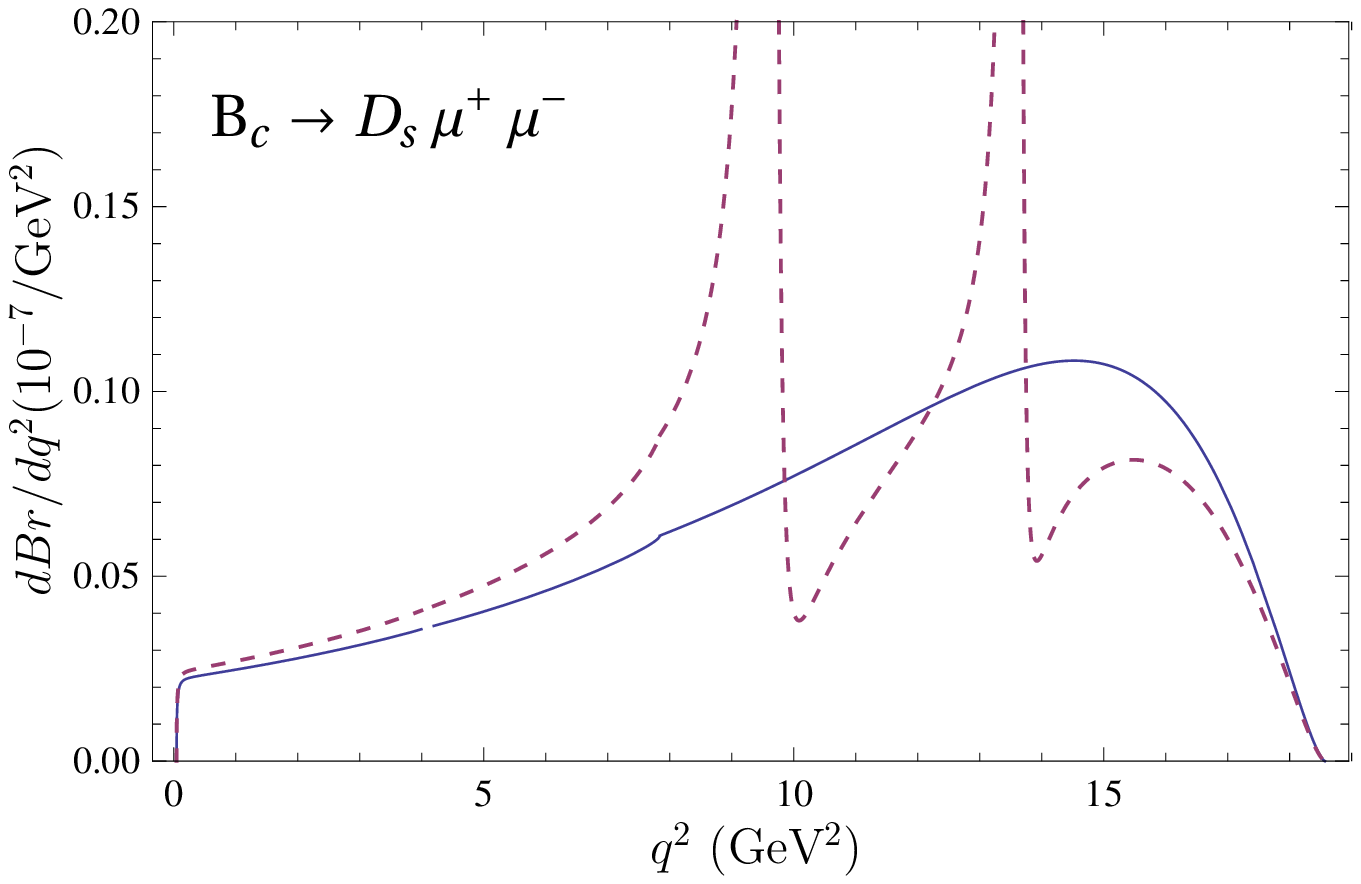}\ \ \ \ \ 
\includegraphics[width=7.5cm]{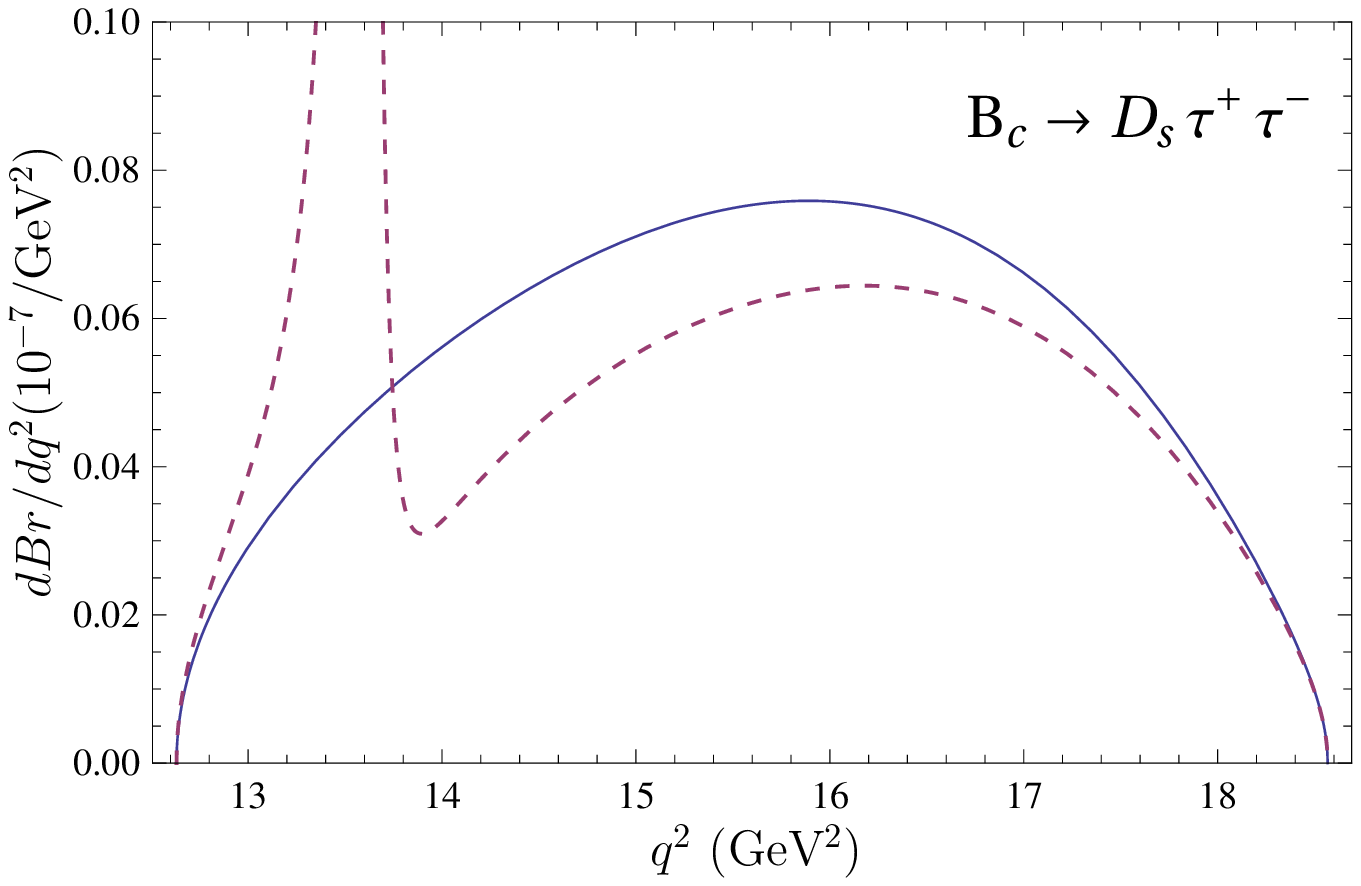}

  \caption{Predictions for the differential $B_c \to
    D_s$ decay branching fractions.  Nonresonant
    and resonant results are plotted by solid and dashed lines,
    respectively.}
  \label{fig:bcd}
\end{figure}

\begin{figure}
  \centering
\includegraphics[width=7.5cm]{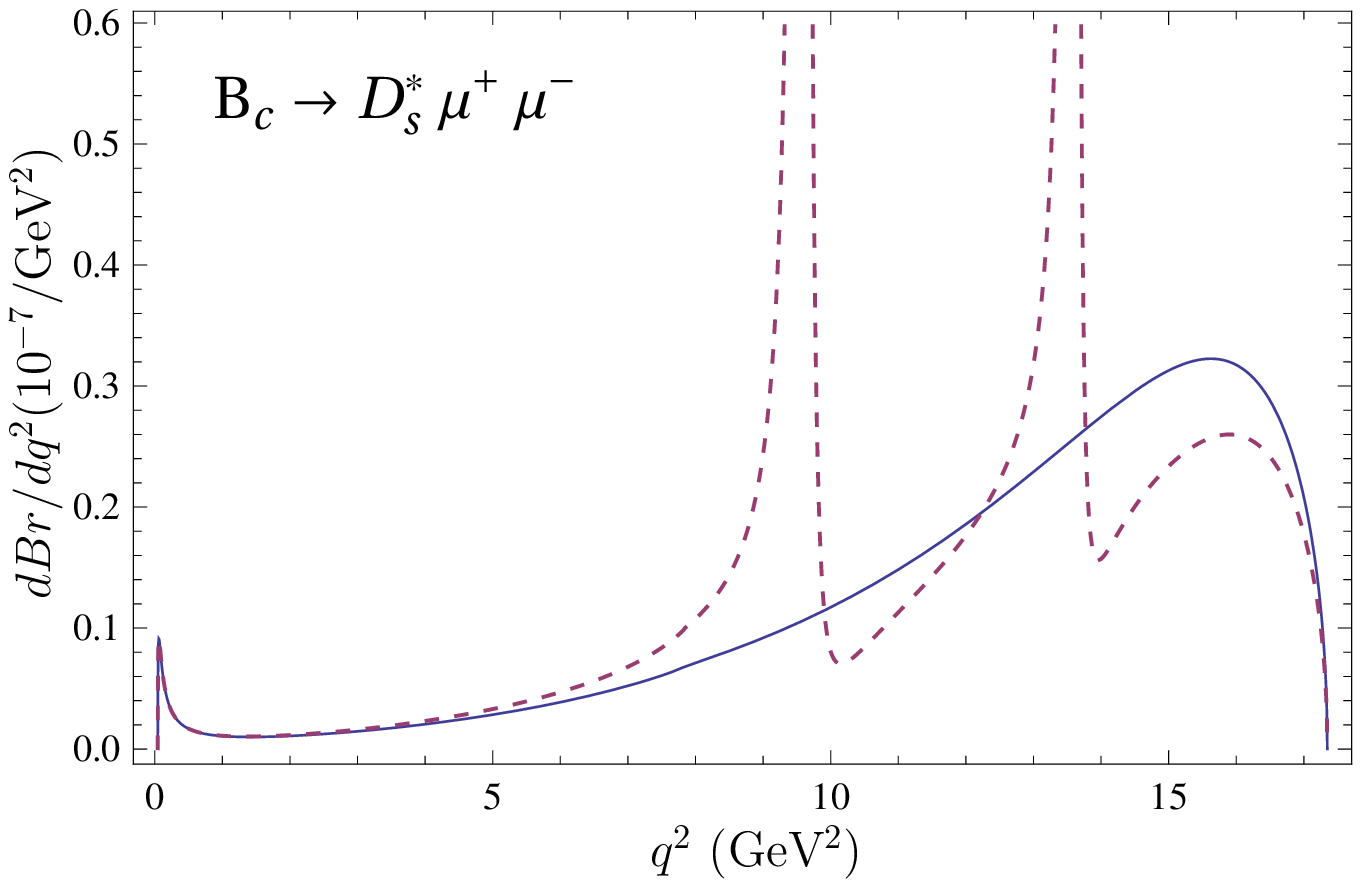}\ \ \ \ \
\includegraphics[width=7.5cm]{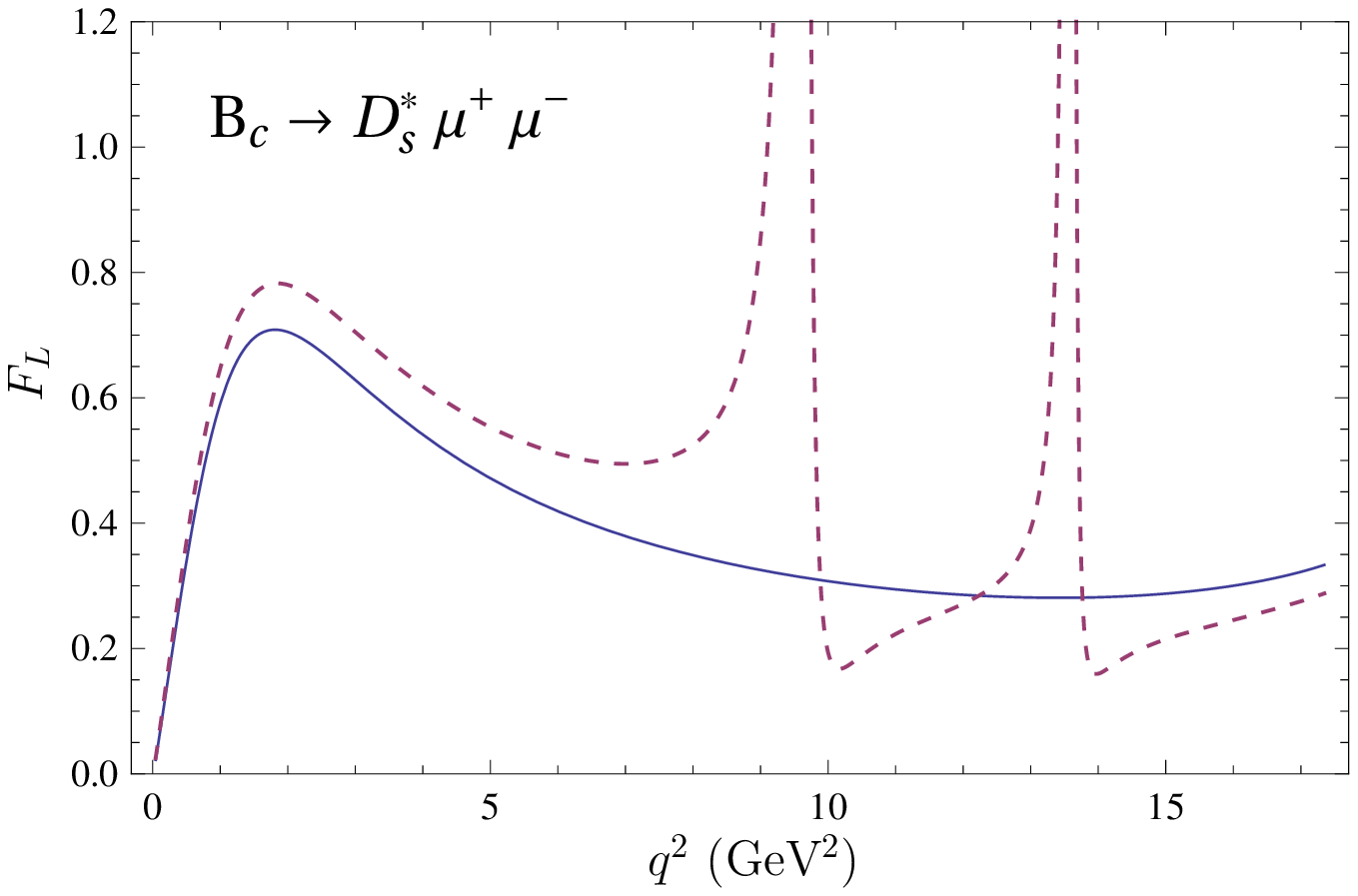}

\vspace*{0.5cm}

\includegraphics[width=7.5cm]{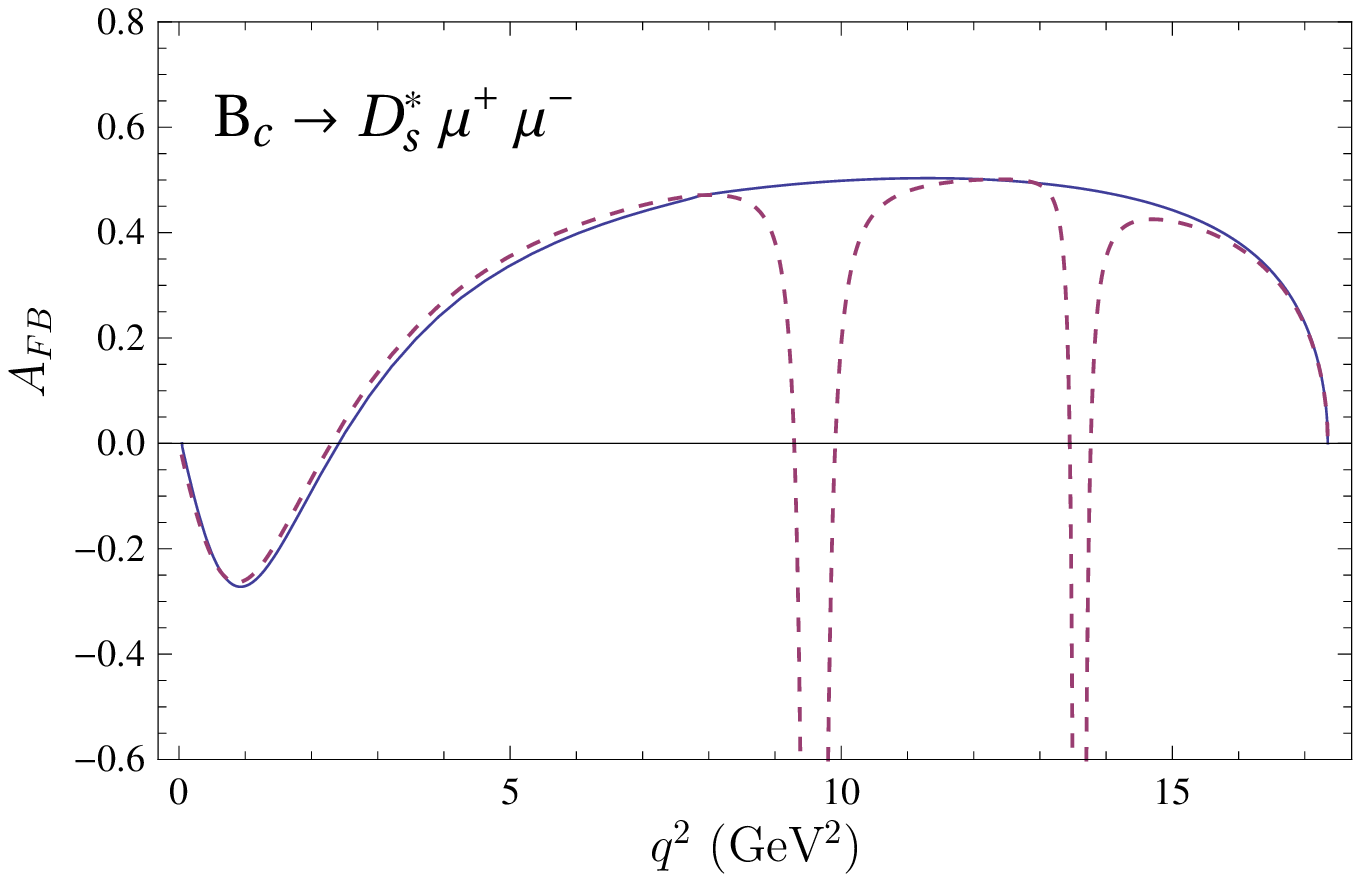}\ \ \ \ \ 
\includegraphics[width=7.5cm]{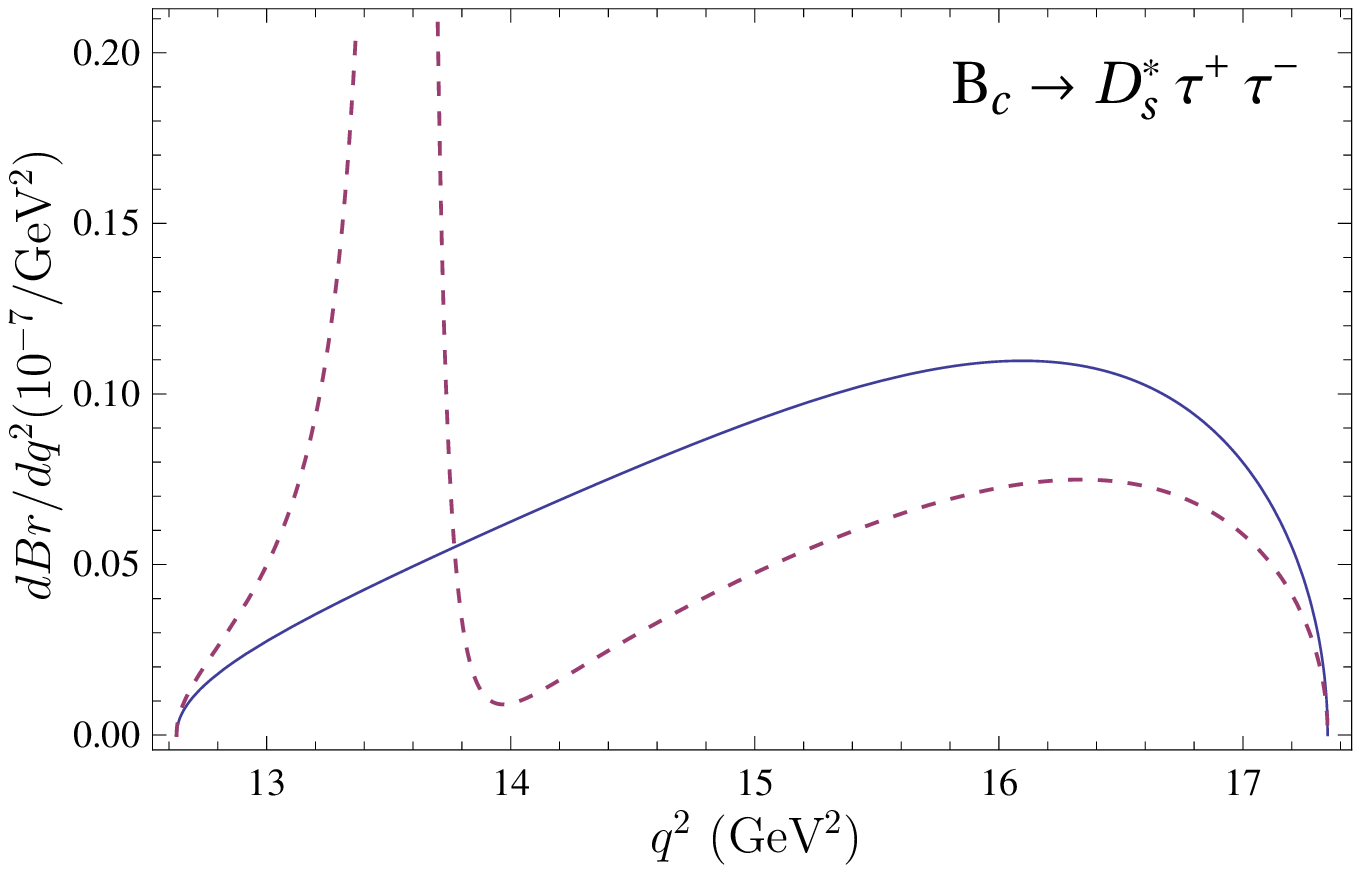}

  \caption{Predictions for the differential
    branching fractions $d Br/d q^2$, the longitudinal polarization
    $F_L$ and muon forward-backward asymmetry $A_{FB}$ for $B_c \to
    D_s^*$ decays. Nonresonant
    and resonant results are plotted by solid and dashed lines,
    respectively.}
  \label{fig:brbcds}
\end{figure}

Next we present our results for the rare semileptonic $B_c$
decays. In Table~\ref{brbc} we compare available theoretical
predictions for the branching fractions of the rare semileptonic $B_c\to
D_s^{(*)}$ and $B_c\to D^{(*)}$ decays. The investigations
\cite{fgikl,ghl,choi}  are based on the relativistic constituent quark
model and light-front quark models, while
the authors of Ref.~\cite{azizi} use three-point QCD sum
rules. Although the results of these approaches are consistent in
the order of magnitude of the branching fractions, 
they differ by more than a factor of 2 for some decay modes. We find the best overall
agreement of our predictions for the branching fractions  with the results
of the relativistic quark model \cite{fgikl}. The differential 
branching fractions, the longitudinal polarization $F_L$ of the final
vector meson and the muon forward-backward asymmetry $A_{FB}$ for the $B_c\to
D_s$ and $B_c\to D_s^*$ transitions in our
model are plotted in Figs.~\ref{fig:bcd} and
\ref{fig:brbcds}. Similar curves have been obtained for $B_c\to
D^{(*)}$ transitions and are not shown here. We
predict the following values of the position of the zero of the
forward-backward asymmetry $A_{FB}$: $q_0^2=2.41$~GeV$^2$ for the $B_c\to
D_s^*\mu^+\mu^-$ decay and $q_0^2=2.46$~GeV$^2$ for the $B_c\to
D^*\mu^+\mu^-$ decay.     

\section{Conclusions}
\label{sec:concl}

In this paper we obtained the form factors of rare
semileptonic decays of the $B$ and $B_c$ mesons in the framework of
the QCD-motivated relativistic quark model based on the 
quasipotential approach. The consideration is done with a
 systematic account of all relativistic effects, which
are very important for such transitions. Particular attention was devoted
to the inclusion of negative-energy contributions and to the
relativistic transformation of the meson wave function from the rest
to the moving reference frame. As a result, the $q^2$ dependence of
these form factors was explicitly determined in the whole accessible
kinematical range without using any ad hoc assumptions and
extrapolations. It is important to point out that the obtained form factors
satisfy all heavy quark and large energy symmetry relations in the
corresponding limits \cite{ffhm}. Note that the resulting decay form factors are expressed through
the overlap integrals of the initial and final meson wave
functions. The relativistic wave
functions, obtained previously in the investigations of the meson mass spectra,
were used for the numerical calculations. This significantly improves
the reliability  of the  
calculated form factors. On the basis of these form factors branching
fractions and different differential decay distributions were obtained.

First we tested our model by confronting its results for the $B\to
K\mu^+\mu^-$ and $B\to K^*\mu^+\mu^-$ decays with the available
detailed experimental data. It was found that the total and
differential branching fractions, the $K^*$ meson longitudinal
polarization fraction $F_L$ and the muon forward-backward asymmetry
$A_{FB}$ agree well with data.

Secondly, we presented detailed predictions for the rare
semileptonic decays of the $B_c$ meson which can be investigated in the
LHCb experiment at CERN, where the $B_c$ mesons are expected to be copiously
produced. Finally, we compare our results on these decays with
the ones previously available in the literature in
Table~\ref{brbc}. The predictions for the differential branching
fractions, the vector meson longitudinal polarization fraction $F_L$ and muon
forward-backward asymmetry $A_{FB}$ are also given in Figs.~\ref{fig:bcd} and
\ref{fig:brbcds}.      

\acknowledgements
The authors are grateful to M. Ivanov, V. Matveev, D. Melikhov,
M. M\"uller-Preussker, N. Nikitin and 
V. Savrin for support and discussions.  
This work was supported in part by  the Deutsche
Forschungsgemeinschaft under contract Eb 139/4-1 and the Russian
Foundation for Basic Research (RFBR) grants
No.08-02-00582 and No.10-02-91339.

\appendix*
\section{Tensor form factors of rare $B$ and $B_c$ decays}

{\it(a) $B\to P$ ($B\to K$, $B_c\to D_s$ and $B_c\to D$) transitions (see Eq.~(\ref{eq:pff2}))}

\begin{equation}
  \label{eq:fft}
  f_T(q^2)=f_T^{(1)}(q^2)+\varepsilon f_T^{S(2)}(q^2)
+(1-\varepsilon) f_T^{V(2)}(q^2),
\end{equation}

\begin{eqnarray}
  \label{eq:ft}
  f_T^{(1)}(q^2)&=&(M_B+M_P)\sqrt{\frac{E_P}{M_B}}\int \frac{{\rm
  d}^3p}{(2\pi)^3} \bar\Psi_P\left({\bf p}+\frac{\epsilon_q}{E_P+M_P}{\bf
  \Delta} \right)\sqrt{\frac{\epsilon_f(p+\Delta)+
  m_f}{2\epsilon_f(p+\Delta)}}\sqrt{\frac{\epsilon_b(p)+
  m_b}{2\epsilon_b(p)}}\cr
&&\times\Biggl\{\frac1{\epsilon_f(p+\Delta)+m_f}+\frac{({\bf
p\Delta})}{{\bf \Delta}^2}\Bigglb(\frac1{\epsilon_f(p+\Delta)+m_f}-\frac1{\epsilon_b(p)+m_b}\Biggrb) \cr
&&+\frac23\frac{{\bf p}^2}{E_P+M_P}
\left(\frac1{\epsilon_f(p+\Delta)+m_f}-\frac1{\epsilon_q(p)+m_q}\right)\cr
&&\times
\left(\frac1{\epsilon_f(p+\Delta)+m_f}+
\frac1{\epsilon_b(p)+m_b}\right)\Biggr\}\Psi_{B}({\bf p}),  
\end{eqnarray}

\begin{eqnarray}
  \label{eq:fts}
  f_T^{S(2)}(q^2)&=&-(M_B+M_P)\sqrt{\frac{E_P}{M_B}}\int \frac{{\rm
  d}^3p}{(2\pi)^3} \bar\Psi_P\left({\bf p}+\frac{2\epsilon_q}{E_P+M_P}{\bf
  \Delta} \right)\sqrt{\frac{\epsilon_f(p+\Delta)+
  m_f}{2\epsilon_f(p+\Delta)}}\cr
&&\times
\Biggl\{\frac1{2\epsilon_f(p+\Delta)(\epsilon_f(p+\Delta)+m_f)}
\left(1+\frac{\epsilon_f(p+\Delta)-m_f}{2m_b}
\frac{({\bf p\Delta})}{{\bf \Delta}^2}\right)\cr
&&\times
\left[M_P-\epsilon_f\left(p+\frac{2\epsilon_q}{E_P+M_P}\Delta\right)
-\epsilon_q\left(p+\frac{2\epsilon_q}{E_P+M_P}\Delta \right)\right]\cr
&&+ \frac{({\bf p\Delta})}{{\bf \Delta}^2}\Biggl[ \frac{M_B+M_f-\epsilon_b(p)-\epsilon_q(p)-\epsilon_f\left(p+\frac{2\epsilon_q}{E_P+M_P}\Delta\right)
-\epsilon_q\left(p+\frac{2\epsilon_q}{E_P+M_P}\Delta
\right)}{2\epsilon_f(p+\Delta)(\epsilon_f(p+\Delta)+m_f)}\cr
&&+\frac{M_B-M_P-\epsilon_b(p)-\epsilon_q(p)+\epsilon_f\left(p+\frac{2\epsilon_q}{E_P+M_P}\Delta\right)
+\epsilon_q\left(p+\frac{2\epsilon_q}{E_P+M_P}\Delta
\right)}{2m_b(\epsilon_b(p+\Delta)+m_b)}\Biggr]\Biggr\}\Psi_B({\bf p}),\cr
&&
\end{eqnarray}

\begin{eqnarray}
  \label{eq:ftv}
  f_T^{V(2)}(q^2)&=&(M_B+M_P)\sqrt{\frac{E_P}{M_B}}\int \frac{{\rm
  d}^3p}{(2\pi)^3} \bar\Psi_P\left({\bf p}+\frac{2\epsilon_q}{E_P+M_P}{\bf
  \Delta} \right)\sqrt{\frac{\epsilon_f(p+\Delta)+
  m_f}{2\epsilon_f(p+\Delta)}}\cr
&&\times
\frac{({\bf p\Delta})}{{\bf \Delta}^2}\Biggl\{
\frac{M_B-\epsilon_b(p)-\epsilon_q(p)}{2\epsilon_f(p+\Delta)(\epsilon_f(p+\Delta)+m_f)}\cr
&&
+\frac1{2m_b}\Biggl(\frac1{\epsilon_b(p+\Delta)+m_b}-
\frac{\epsilon_f(p+\Delta)-m_f}{2\epsilon_f(p+\Delta)(\epsilon_f(p+\Delta)+m_f)}\Biggr)\cr
&&\times
\left[M_P-\epsilon_f\left(p+\frac{2\epsilon_q}{E_P+M_P}\Delta\right)
-\epsilon_q\left(p+\frac{2\epsilon_q}{E_P+M_P}\Delta
\right)\right]\Biggr\}\Psi_B({\bf p}),
\end{eqnarray}
where the superscripts ``(1)" and ``(2)" correspond to Figs.~\ref{d1} and
\ref{d2}, $\varepsilon$ is the mixing coefficient in the confinement
potential indicated in Eq.~(\ref{vlin}) and
\[\Delta\equiv \left|{\bf \Delta}\right|=\sqrt{\frac{(M_{B}^2+M_P^2-q^2)^2}
{4M_{B}^2}-M_P^2},\]
\[ E_P=\sqrt{M_P^2+{\bf \Delta}^2}, \quad
 \epsilon_Q(p+a
\Delta)=\sqrt{m_Q^2+({\bf p}+a{\bf \Delta})^2} \quad
(Q=b,c,s,u,d). \]
Here $B$ stands for the $B$ or $B_c$ meson, $P=K,D_s,D$ is the final
pseudoscalar meson, $f=s,d$ is the final active quark and $q=u,d,c$
denotes the corresponding spectator quark.

  \bigskip
{\it(b) $B\to V$ ($B\to K^*$, $B_c\to D_s^*$ and $B_c\to D^*$)  transitions (see
Eqs.~(\ref{eq:vff3}) and (\ref{eq:vff4}))}
\nopagebreak
\begin{equation}
  \label{eq:ft1}
  T_1(q^2)=T_1^{(1)}(q^2)+\varepsilon T_1^{S(2)}(q^2)
+(1-\varepsilon) T_1^{V(2)}(q^2),
\end{equation}

\begin{eqnarray}
  \label{eq:t1}
T_1^{(1)}(q^2)&=&\sqrt{\frac{E_V}{ M_B}}\int \frac{d^3p}{(2\pi)^3}
\bar \Psi_V\left({\bf p}+\frac{2\epsilon_q}{
E_V+M_V}{\bf\Delta}\right) \sqrt{\frac{\epsilon_f(p+\Delta) +m_f}
{ 2\epsilon_f(p+\Delta)}}\sqrt{\frac{\epsilon_b (p)+m_b}{
2\epsilon_b(p)}} \cr
& &\times\Biggl\{ 1+\frac{{\bf p}^2/3+({\bf p\Delta})}
{(\epsilon_f(p+\Delta)+ m_f) (\epsilon_b(p)+m_b)}+\cr
&& \frac{{\bf
    p}^2{\bf\Delta}^2}{3(E_V+M_V)(\epsilon_q(p)+m_q)(\epsilon_f
(p+\Delta)+ m_f) (\epsilon_b(p)+m_b)} \cr
& & +(M_B-E_V)\Biggl[\frac{1}
{\epsilon_f(p+\Delta)+m_f}+\frac{({\bf p\Delta})}{{\bf
\Delta}^2}\left(\frac{1}{\epsilon_f(p+\Delta)+m_f}+
\frac{1}{\epsilon_b(p)+m_b}\right)\nonumber\\
& &-\frac{{\bf p}^2}{ 3(E_V+M_V)}\Biggl(\frac{1}{
\epsilon_q(p)+m_q}\left(\frac{1}{ \epsilon_b(p)+m_b}
-\frac{1}{ \epsilon_f(p+\Delta)+m_f}\right)\nonumber\\
& &+\frac{2}{ (\epsilon_f(p) +m_f)^2}
\Biggr)\Biggr]\Biggr\}\Psi_B({\bf p}),
\end{eqnarray}

\begin{eqnarray}
  \label{eq:t1s}
T_1^{S(2)}(q^2)&=&\sqrt{\frac{E_V}{ M_B}} \int\frac{d^3p}{(2\pi)^3}
\bar\Psi_V\left({\bf p}+\frac{2\epsilon_q}{
E_V+M_V}{\bf\Delta}\right) \sqrt\frac{\epsilon_f(p+\Delta)+m_f}{
2\epsilon_f(p+\Delta)} \cr
& &\times \Biggl\{\frac{\epsilon_f(p+\Delta)-
m_f}{2\epsilon_f(p+\Delta)
(\epsilon_f(p+\Delta)+m_f)}\left(1+\frac{M_B-E_V}{ 2m_b}\frac{({\bf
p\Delta})}{{\bf \Delta}^2}\right) \cr
& &\times\left(M_V-\epsilon_f\left(p+\frac{2\epsilon_q}{
E_V+M_V}\Delta\right)- \epsilon_q\left(p+\frac{2\epsilon_q}{
E_V+M_V}\Delta\right)\right)-(M_B-E_V)\frac{({\bf p\Delta})}{
{\bf\Delta}^2} \cr
& &\times\Biggl(\frac{M_B+M_V-\epsilon_b(p)-\epsilon_q(p)-
\epsilon_f\left(p+\frac{2\epsilon_q}{E_V+M_V}\Delta\right)
-\epsilon_q\left(p+\frac{2\epsilon_q}{E_V+M_V}\Delta\right)}{
2m_b(\epsilon_b(p+\Delta)+m_b)} \cr
& &+\frac{M_B-M_V-\epsilon_b(p)-\epsilon_q(p)+
\epsilon_f\left(p+\frac{2\epsilon _q}{E_V+M_V}
\Delta\right)+\epsilon_q\left(p+\frac{2\epsilon_q}{
E_V+M_V}\Delta\right) }{2\epsilon_f(p+\Delta)(\epsilon_f(p+\Delta)
+m_f)}\Biggr)\Biggr\} \Psi_B({\bf p}),\cr
&&
\end{eqnarray}

\begin{eqnarray}
  \label{eq:t1v}
T_1^{V(2)}(q^2)&=&\sqrt{\frac{E_V}{ M_V}}\int\frac{d^3p}{ (2\pi)^3}
\bar\Psi_V\left({\bf p}+\frac{2\epsilon_q}{
E_V+M_V}{\bf\Delta}\right) \sqrt{\frac{\epsilon_f(p+\Delta)+m_f}{
2\epsilon_f(p+\Delta)}}\nonumber\\
& &\times \Biggl\{\frac{\epsilon_f(p+\Delta)-
m_f}{2\epsilon_f(p+\Delta)
(\epsilon_f(p+\Delta)+m_f)}\left(1+\frac{M_B-E_V}{ 2m_b}\frac{({\bf
p\Delta})}{{\bf\Delta}^2}\right)\nonumber\\
& &\times\left(M_V-\epsilon_f\left(p+\frac{2\epsilon_q}{
E_V+M_V}\Delta\right)- \epsilon_q\left(p+\frac{2\epsilon_q}{
E_V+M_V}\Delta\right)\right)+(M_B-E_V)\frac{({\bf p\Delta})}{
{\bf\Delta}^2}\nonumber\\
& &\times\Biggl(\frac{M_V-\epsilon_f\left(p+\frac{2\epsilon_q}{
E_V+M_V}\Delta\right) -\epsilon_q\left(p+\frac{2\epsilon_q}{
E_V+M_V}\Delta\right)}{ 2m_b(\epsilon_b(p+\Delta)+m_b)}\nonumber\\
& &
+\frac{M_B-\epsilon_b(p)-\epsilon_q(p)}{
2\epsilon_f(p+\Delta) (\epsilon_f(p+\Delta)+m_f)}\Biggr)\Biggr\}\Psi_B({\bf p}),
\end{eqnarray}

\begin{equation}
  \label{eq:ft2}
  T_2(q^2)=T_2^{(1)}(q^2)+\varepsilon T_2^{S(2)}(q^2)
+(1-\varepsilon) T_2^{V(2)}(q^2),
\end{equation}

\begin{eqnarray}
  \label{eq:t2}
T_2^{(1)}(q^2)&=&\frac{2\sqrt{E_V M_B}}{M_B^2-M_V^2}\int \frac{d^3p}{(2\pi)^3}
\bar \Psi_V\left({\bf p}+\frac{2\epsilon_q}{
E_V+M_V}{\bf\Delta}\right) \sqrt{\frac{\epsilon_f(p+\Delta) +m_f}
{ 2\epsilon_f(p+\Delta)}}\sqrt{\frac{\epsilon_b (p)+m_b}{
2\epsilon_b(p)}} \cr
& &\times\Biggl\{ (M_B-E_V)\Biggl[1+\frac{{\bf p}^2/3+({\bf p\Delta})}
{(\epsilon_f(p+\Delta)+ m_f) (\epsilon_b(p)+m_b)}+\cr
&& \frac{{\bf
    p}^2{\bf\Delta}^2}{3(E_V+M_V)(\epsilon_q(p)+m_q)(\epsilon_f
(p+\Delta)+ m_f) (\epsilon_b(p)+m_b)}\Biggr] \cr
& & +{\bf\Delta}^2\Biggl[\frac{1}
{\epsilon_f(p+\Delta)+m_f}+\frac{({\bf p\Delta})}{{\bf
\Delta}^2}\left(\frac{1}{\epsilon_f(p+\Delta)+m_f}+
\frac{1}{\epsilon_b(p)+m_b}\right)\nonumber\\
& &-\frac{{\bf p}^2}{ 3(E_V+M_V)}\Biggl(\frac{1}{
\epsilon_q(p)+m_q}\left(\frac{1}{ \epsilon_b(p)+m_b}
-\frac{1}{ \epsilon_f(p+\Delta)+m_f}\right)\nonumber\\
& &+\frac{2}{ (\epsilon_f(p) +m_f)^2}
\Biggr)\Biggr]\Biggr\}\Psi_B({\bf p}),
\end{eqnarray}

\begin{eqnarray}
  \label{eq:t2s}
T_2^{S(2)}(q^2)&=&\frac{2\sqrt{E_V M_B}}{M_B^2-M_V^2} \int\frac{d^3p}{(2\pi)^3}
\bar\Psi_V\left({\bf p}+\frac{2\epsilon_q}{
E_V+M_V}{\bf\Delta}\right) \sqrt\frac{\epsilon_f(p+\Delta)+m_f}{
2\epsilon_f(p+\Delta)} \cr
& &\times \Biggl\{\frac{\epsilon_f(p+\Delta)-
m_f}{2\epsilon_f(p+\Delta)
(\epsilon_f(p+\Delta)+m_f)}\left(M_B-E_V+\frac{({\bf
p\Delta})}{ 2m_b}\right) \cr
& &\times\left(M_V-\epsilon_f\left(p+\frac{2\epsilon_q}{
E_V+M_V}\Delta\right)- \epsilon_q\left(p+\frac{2\epsilon_q}{
E_V+M_V}\Delta\right)\right)\cr
& &-({\bf p\Delta})\Biggl(\frac{M_B+M_V-\epsilon_b(p)-\epsilon_q(p)-
\epsilon_f\left(p+\frac{2\epsilon_q}{E_V+M_V}\Delta\right)
-\epsilon_q\left(p+\frac{2\epsilon_q}{E_V+M_V}\Delta\right)}{
2m_b(\epsilon_b(p+\Delta)+m_b)} \cr
& &+\frac{M_B-M_V-\epsilon_b(p)-\epsilon_q(p)+
\epsilon_f\left(p+\frac{2\epsilon _q}{E_V+M_V}
\Delta\right)+\epsilon_q\left(p+\frac{2\epsilon_q}{
E_V+M_V}\Delta\right) }{2\epsilon_f(p+\Delta)(\epsilon_f(p+\Delta)
+m_f)}\Biggr)\Biggr\} \Psi_B({\bf p}),\cr
&&
\end{eqnarray}

\begin{eqnarray}
  \label{eq:t2v}
T_2^{V(2)}(q^2)&=&\frac{2\sqrt{E_V M_B}}{M_B^2-M_V^2} \int\frac{d^3p}{ (2\pi)^3}
\bar\Psi_V\left({\bf p}+\frac{2\epsilon_q}{
E_V+M_V}{\bf\Delta}\right) \sqrt{\frac{\epsilon_f(p+\Delta)+m_f}{
2\epsilon_f(p+\Delta)}} \cr
& &\times \Biggl\{\frac{\epsilon_f(p+\Delta)-
m_f}{2\epsilon_f(p+\Delta)
(\epsilon_f(p+\Delta)+m_f)}\left(M_B-E_V+\frac{({\bf
p\Delta})}{ 2m_b}\right) \cr
& &\times\left(M_V-\epsilon_f\left(p+\frac{2\epsilon_q}{
E_V+M_V}\Delta\right)- \epsilon_q\left(p+\frac{2\epsilon_q}{
E_V+M_V}\Delta\right)\right) \cr
& &+({\bf p\Delta})\Biggl(\frac{M_V-\epsilon_f\left(p+\frac{2\epsilon_q}{
E_V+M_V}\Delta\right) -\epsilon_q\left(p+\frac{2\epsilon_q}{
E_V+M_V}\Delta\right)}{ 2m_b(\epsilon_b(p+\Delta)+m_b)} \cr
& &
+\frac{M_B-\epsilon_b(p)-\epsilon_q(p)}{
2\epsilon_f(p+\Delta) (\epsilon_f(p+\Delta)+m_f)}\Biggr)\Biggr\}\Psi_B({\bf p}),
\end{eqnarray}

\begin{equation}
  \label{eq:ft3}
  T_3(q^2)=T_3^{(1)}(q^2)+\varepsilon T_3^{S(2)}(q^2)
+(1-\varepsilon) T_3^{V(2)}(q^2),
\end{equation}

\begin{eqnarray}
  \label{eq:t3}
T_3^{(1)}(q^2)&=&\sqrt{\frac{E_V}{ M_B}}\int \frac{d^3p}{(2\pi)^3}
\bar \Psi_V\left({\bf p}+\frac{2\epsilon_q}{
E_V+M_V}{\bf\Delta}\right) \sqrt{\frac{\epsilon_f(p+\Delta) +m_f}
{ 2\epsilon_f(p+\Delta)}}\sqrt{\frac{\epsilon_b (p)+m_b}{
2\epsilon_b(p)}} \cr
& &\times\Biggl\{-\Biggl( 1+\frac{{\bf p}^2/3-({\bf p\Delta})}
{(\epsilon_f(p+\Delta)+ m_f) (\epsilon_b(p)+m_b)}+\cr
&& \frac{2{\bf
    p}^2{\bf\Delta}^2}{3(E_V+M_V)(\epsilon_f(p+\Delta)+ m_f)^2
  (\epsilon_b(p)+m_b)} \Biggr) +\frac{M_BE_V+M_V^2}{M_B}\Biggl[\frac{1}
{\epsilon_f(p+\Delta)+m_f}\cr
& &+\frac{({\bf p\Delta})}{{\bf
\Delta}^2}\left(\frac{1}{\epsilon_f(p+\Delta)+m_f}+
\frac{1}{\epsilon_b(p)+m_b}+\frac{2(M_B-E_V)}{(\epsilon_f(p+\Delta)+
  m_f) (\epsilon_b(p)+m_b)}\right) \cr
& &-\frac{{\bf p}^2}{ 3(E_V+M_V)}\Biggl(\frac{1}{
\epsilon_q(p)+m_q}\left(\frac{1}{ \epsilon_b(p)+m_b}
-\frac{1}{ \epsilon_f(p+\Delta)+m_f}\right)\nonumber\\
& &+\frac{2}{ (\epsilon_f(p) +m_f)^2}
+\frac{M_B-E_V}{(\epsilon_f(p+\Delta)+
  m_f) (\epsilon_b(p)+m_b)}\cr
&&\times\left(\frac{2}{ \epsilon_f(p+\Delta)+m_f }
-\frac{1}{\epsilon_q(p)+m_q}\right)\Biggr)\Biggr]\Biggr\}\Psi_B({\bf p}),
\end{eqnarray}

\begin{eqnarray}
  \label{eq:t3s}
T_3^{S(2)}(q^2)&=&\sqrt{\frac{E_V}{ M_B}} \int\frac{d^3p}{(2\pi)^3}
\bar\Psi_V\left({\bf p}+\frac{2\epsilon_q}{
E_V+M_V}{\bf\Delta}\right) \sqrt\frac{\epsilon_f(p+\Delta)+m_f}{
2\epsilon_f(p+\Delta)} \cr
& &\times \Biggl\{\frac{\epsilon_f(p+\Delta)-
m_f}{2\epsilon_f(p+\Delta)
(\epsilon_f(p+\Delta)+m_f)}\left(\frac{M_BE_V+M_V^2}{ 2M_Bm_b}\frac{({\bf
p\Delta})}{{\bf \Delta}^2}-1\right) \cr
& &\times\left(M_V-\epsilon_f\left(p+\frac{2\epsilon_q}{
E_V+M_V}\Delta\right)- \epsilon_q\left(p+\frac{2\epsilon_q}{
E_V+M_V}\Delta\right)\right)-\frac{M_BE_V+M_V^2}{M_B}\frac{({\bf p\Delta})}{
{\bf\Delta}^2} \cr
& &\times\Biggl(\frac{M_B+M_V-\epsilon_b(p)-\epsilon_q(p)-
\epsilon_f\left(p+\frac{2\epsilon_q}{E_V+M_V}\Delta\right)
-\epsilon_q\left(p+\frac{2\epsilon_q}{E_V+M_V}\Delta\right)}{
2m_b(\epsilon_b(p+\Delta)+m_b)} \cr
& &+\frac{M_B-M_V-\epsilon_b(p)-\epsilon_q(p)+
\epsilon_f\left(p+\frac{2\epsilon _q}{E_V+M_V}
\Delta\right)+\epsilon_q\left(p+\frac{2\epsilon_q}{
E_V+M_V}\Delta\right) }{2\epsilon_f(p+\Delta)(\epsilon_f(p+\Delta)
+m_f)}\Biggr)\Biggr\} \Psi_B({\bf p}),\cr
&&
\end{eqnarray}

\begin{eqnarray}
  \label{eq:t3v}
T_3^{V(2)}(q^2)&=&\sqrt{\frac{E_V}{ M_V}}\int\frac{d^3p}{ (2\pi)^3}
\bar\Psi_V\left({\bf p}+\frac{2\epsilon_q}{
E_V+M_V}{\bf\Delta}\right) \sqrt{\frac{\epsilon_f(p+\Delta)+m_f}{
2\epsilon_f(p+\Delta)}}\cr
& &\times \Biggl\{\frac{\epsilon_f(p+\Delta)-
m_f}{2\epsilon_f(p+\Delta)
(\epsilon_f(p+\Delta)+m_f)}\left(\frac{M_BE_V+M_V^2}{ 2M_Bm_b}\frac{({\bf
p\Delta})}{{\bf\Delta}^2}-1\right)\cr
& &\times\left(M_V-\epsilon_f\left(p+\frac{2\epsilon_q}{
E_V+M_V}\Delta\right)- \epsilon_q\left(p+\frac{2\epsilon_q}{
E_V+M_V}\Delta\right)\right)\cr
& &+\frac{M_BE_V+M_V^2}{M_B}\frac{({\bf p\Delta})}{
{\bf\Delta}^2}\Biggl(\frac{M_V-\epsilon_f\left(p+\frac{2\epsilon_q}{
E_V+M_V}\Delta\right) -\epsilon_q\left(p+\frac{2\epsilon_q}{
E_V+M_V}\Delta\right)}{ 2m_b(\epsilon_b(p+\Delta)+m_b)}\cr
& &
+\frac{M_B-\epsilon_b(p)-\epsilon_q(p)}{
2\epsilon_f(p+\Delta) (\epsilon_f(p+\Delta)+m_f)}\Biggr)\Biggr\}\Psi_B({\bf p}),
\end{eqnarray}
where $V=K^*,D_s^*,D^*$ and
\[ \Delta\equiv\left|{\bf \Delta}\right|=\sqrt{\frac{(M_{B}^2+M_V^2-q^2)^2}
{4M_{B}^2}-M_V^2},\qquad
 E_V=\sqrt{M_V^2+{\bf \Delta}^2}. \]

\end{document}